\documentclass[fleqn]{article}
\usepackage{amsmath,amsthm,amssymb,bm,lscape}
\usepackage[body={17cm,25cm}]{geometry}

\newcounter{app}
\newcommand{\app}[1]{
\refstepcounter{app}{\vspace{7mm}
\noindent\Large\bf Appendix
\theapp.
 \ #1 \par \vspace{5mm}}
\setcounter{equation}{0}
\def\theequation{\Alph{app}.\arabic{equation}}}
\begin{document}

\title{The  $TQ$ equation of the 8 vertex model for complex 
elliptic roots of unity}
\author{Klaus Fabricius
\footnote{e-mail Fabricius@theorie.physik.uni-wuppertal.de}\\
 Physics Department, University of Wuppertal, 
42097 Wuppertal, Germany\\ \\
Barry M.~~McCoy\footnote{e-mail mccoy@max2.physics.sunysb.edu}\\
Institute for Theoretical Physics, State University of New York,
Stony Brook,  NY 11794-384}
\date{\today}

\maketitle   

\begin{abstract}
We extend  our studies of the $TQ$ equation
introduced by Baxter in his 1972 solution of the 8 vertex model
with parameter $\eta$ given by $2L\eta=2m_1K+im_2K'$ from $m_2=0$ to
the more general case of complex $\eta.$ We find that there are
several different cases depending on the parity of $m_1$ and $m_2$.

\end{abstract}

\section{Introduction}

In 1972 Baxter published one of the most important and influential
papers in statistical mechanics of the $20^{th}$ century \cite{bax72}.
A loose statement is that in this paper Baxter ``solves the 8 vertex model''.  
A  more precise statement is that Baxter demonstrates the existence of
a matrix $Q(v)$ which satisfies the following functional equation with the
transfer matrix $T(v)$ of the 8 vertex model
\begin{equation}
T(v)Q(v)=[h(v+\eta)]^NQ(v-2\eta)+[h(v-\eta)]^NQ(v+2\eta)
\label{tq1}
\end{equation}
with
\begin{eqnarray}
&&[T(v),Q(v')]=0\label{tq2}\\
&&[Q(v),Q(v')]=0.\label{tq3}
\end{eqnarray}
Here $N$ is the number of horizontal sites, $h(v)$ is a suitable 
quasiperiodic function, and $\eta$ is a  constant that is present in the
Boltzmann weights of the model which satisfies
\begin{equation}
2L\eta=2m_1K+im_2K'
\label{tq4}
\end{equation}
where $L,~ m_1$ and $m_2$ are integers,  $K$ is the complete elliptic 
integral of the first kind 
of modulus $k$ and $K'$ is the complete elliptic integral of the
complementary modulus
$k'=(1-k^2)^{1/2}.$
The values of $\eta$ satisfying the condition (\ref{tq4}) are 
referred to in the physics literature as elliptic roots of unity
(which we will abbreviate to ``roots of unity''). In the mathematics literature
these values of $\eta$ are referred to as division points.

In the past 35 years there have been many papers which extend,
generalize, and comment on equations (\ref{tq1})-(\ref{tq3}) which
are often collectively referred to as ``the TQ equation''. Some of this
literature is by Baxter \cite{bax731}-\cite{bax02}, Takhtadzhan 
and Faddeev \cite{tf}, Zabrodin \cite{Zab}, Deguchi \cite{deg1,deg2}, 
Bazhanov and Mangazeev \cite{baz1}-\cite{baz3},
 Roan \cite{roan}  and the present
authors \cite{fm1}-\cite{klaus}. Each paper has added to
our understanding of the original paper \cite{bax72} but each paper
leaves some questions unexplored. 

The transfer matrix $T(v)$ of the 8 vertex model has two discrete
symmetries expressed by the commutation relations
\begin{equation}
[T(v),S]=0~~~~~ [T(v),R]=0\label{rcomm}
\end{equation}
where (in the canonical basis given in sec. 2)
\begin{equation}
S=\sigma^z_1\otimes
\sigma^z_2\otimes \cdots \otimes\sigma^z_N~~~~~~~ 
R=\sigma^x_1\otimes \sigma^x_2 \otimes \cdots\otimes
\sigma^x_N\label{rdef}
\end{equation}
We note that
\begin{equation}
RS=(-1)^NSR
\end{equation}
and  from (\ref{rcomm}) it follows that
\begin{equation}
[T(v),RS]=0 \label{rscomm}
\end{equation}

If the transfer matrix $T(v)$ were nondegenerate then  the matrix
$Q(v)$ of (\ref{tq1})-(\ref{tq3}) would also satisfy 
(\ref{rcomm}) and (\ref{rscomm}). However, when (\ref{tq4}) holds 
the matrix $T(v)$ has degenerate eigenvalues and the matrix $Q(v)$ no
longer needs to satisfy these commutation relations. For this reason
the matrix $Q$ which satisfies (\ref{tq1})-(\ref{tq3}) is not
uniquely defined when (\ref{tq4}) holds. 

The $Q(v)$ matrix can be chosen to have the same discrete symmetries
(\ref{rcomm})  and (\ref{rscomm}) as the transfer
matrix. This choice is
made in the 1973 paper of Baxter \cite{bax731}. We call this choice 
$Q_{73}(v)$. This matrix $Q_{73}(v)$ is defined for all values of $\eta.$ 
In the original paper \cite{bax72} of 1972 the condition
(\ref{tq4}) is imposed and a matrix $Q(v)$ is constructed which
is nondegenerate. We call this matrix $Q_{72}(v)$. 

We have investigated this construction of $Q_{72}(v)$ in 
the series of papers \cite{fm1}-\cite{klaus}. In the first paper 
of the series \cite{fm1} we found for $m_2=0$ and $m_1$ odd  
that the construction of ref. \cite{bax72} 
gives a $Q_{72}(v)$ matrix which commutes with $S$ but 
does not commute with $R$ and $RS$.
We also found that the construction fails 
for $L$ odd and $m_1$ even. For this case a new construction was found
in  \cite{klaus} which contains a parameter $t$.
We call $Q_{72}^{(1)}(v)$ the $Q(v)$ matrix constructed by means
of \cite{bax72} and $Q^{(2)}_{72}(v;t)$ the matrix constructed by use of
the procedure of \cite{klaus}.

In this paper we extend these constructions by considering the
general case of (\ref{tq4}) with $m_2\neq 0$. We find that there are
four distinct cases depending on the parity of $m_1$ and $m_2.$ When necessary
we denote these cases as $Q^{(1)}_{72oe}(v),
Q^{(1)}_{72oo}(v),Q^{(1)}_{72eo}(v),$ and $Q^{(2)}_{72ee}(v;t)$ 
where the first (second)
subscript indicates the parity of $m_1~ (m_2)$.
These cases are distinguished by different commutation relations with
the operators $S,R$ and $RS$ as follows.
 
\vspace{.1in}

{\bf Case 1} $m_1$ odd, $m_2$ even, N unrestricted
\begin{equation}
[Q_{72oe}^{(1)}(v),S]=0,~[Q_{72oe}^{(1)}(v),R]\neq
0,~[Q_{72oe}^{(1)}(v),RS]\neq 0
\label{c1}
\end{equation}

{\bf Case 2} $m_1$ odd, $m_2$ odd, N unrestricted
\begin{equation}
[Q_{72oo}^{(1)}(v),S]\neq0,~[Q_{72oo}^{(1)}(v),R]\neq 0,
~[Q_{72oo}^{(1)}(v),RS]=0
\label{c3}
\end{equation}

{\bf Case 3} $m_1$ even, $m_2$ odd, N unrestricted
\begin{equation}
[Q_{72eo}^{(1)}(v),S]\neq0,~[Q_{72eo}^{(1)}(v),R]=
0,~[Q_{72eo}^{(1)}(v),RS]\neq 0
\label{c4}
\end{equation}

\vspace{.1in}

{\bf Case 4} $m_1$ even, $m_2$ even, $N$ even. 

We find that there are matrices
$Q_{72ee}^{(2)}(v;t)$ for $t=n\eta$ and $t=(n+1/2)\eta$ with $n$ an
integer and thus there are
several subcases to be distinguished

\vspace{.1in}

{\bf Case 4A} $t=n\eta$
\begin{equation}
[Q_{72ee}^{(2)}(v;n\eta),S]= 0,~[Q_{72ee}^{(2)}(v;n\eta),R]\neq
0,~[Q_{72ee}^{(2)}(v;n\eta),RS]\neq 0
\label{c4a}
\end{equation}

{\bf Case 4B} $t=(n+1/2)\eta$, 
$m_1\equiv 0 ~({\rm mod}4)$, and $m_2\equiv 0 ~({\rm mod}4)$
\begin{equation}
[Q_{72ee}^{(2)}(v;(n+1/2)\eta),S]= 0,~[Q_{72ee}^{(2)}(v;(n+1/2)\eta),R]\neq
0,~[Q_{72ee}^{(2)}(v;(n+1/2)\eta),RS]\neq 0
\label{c4b}
\end{equation}
The matrix $Q^{(2)}_{72ee}(v;n\eta)$ is similar to the matrix
 $Q^{(2)}_{72ee}(v;(n'+1/2)\eta).$\\\\

{\bf Case 4C} $t=(n+1/2)\eta$,
$m_1\equiv 2~~({\rm mod} 4),~~m_2\equiv 2~~({\rm mod} 4)$ 
\begin{equation}
[Q_{72ee}^{(2)}(v;(n+1/2)\eta),S]\neq 0,~[Q_{72ee}^{(2)}(v;(n+1/2)\eta),R]=
0,~[Q_{72ee}^{(2)}(v;(n+1/2)\eta),RS]\neq 0
\label{c4c}
\end{equation}

{\bf Case 4D} $t=(n+1/2)\eta$,
$m_1\equiv 0~~({\rm mod} 4),~~m_2\equiv 2~~({\rm mod} 4)$ 
\begin{equation}
[Q_{72ee}^{(2)}(v;(n+1/2)\eta),S]\neq 0,
~[Q_{72ee}^{(2)}(v;(n+1/2)\eta),R]\neq 0,
[Q_{72ee}^{(2)}(v;(n+1/2)\eta),RS]= 0
\label{c4d}
\end{equation}

No matrix $Q^{(2)}_{72ee}(v,(n+1/2)\eta)$ exists for $m_1\equiv 2~({\rm
mod}4),~~m_2\equiv 0~({\rm mod}4)$.
In addition it is known from numerical computations \cite{klaus} that 
for general
values of $t$ a matrix can be constructed that satisfies
(\ref{tq1}) and (\ref{tq2}) but not (\ref{tq3}) which does not commute
with any of $S,R$ or $RS$.

There remains one case for which no matrix  has yet been found by use
of the methods of refs. \cite{bax72} or \cite{klaus} which satisfies
(\ref{tq1})-(\ref{tq3}); this is $m_1$ even, $m_2$ even and $N$
odd. It has been seen in \cite{fm2} from numerical computations that a
TQ equation for eigenvalues holds and the eigenvalues of $Q(v)$ have
unique properties not seen in cases 1-4. The particular case
$\eta=2K/3$ is extensivly treated in \cite{baz1} where  unique properties also
exist for the 6 vertex limit \cite{strog1}-\cite{murray}.

In sec. 2 we review the formalism of the 8 vertex model for the case
that (\ref{tq4}) holds with $m_2\neq 0.$ This involves a modification
of the theta functions $\Theta(v)$ and $H(v)$ which was first
introduced in \cite{bax731}. The properties of these modified theta
functions are summarized in  appendix A where we also prove various
identities which will be used in the text..
In sec. 3 we review the three steps of the  construction of the 
$Q_{72}(v)$ matrix
of ref. \cite{bax72} which uses the auxiliary matrices $Q_R(v)$ and
$Q_L(v).$ The explicit construction of $Q_R(v)$ is given in sec. 4
 with special attention to the recent discovery
\cite{klaus} that the principles of this construction lead to two 
different $Q$ matrices $Q^{(1)}_{72}(v)$ and $Q^{(2)}_{72}(v;t)$.
 In sec. 5 we use the methods of ref.\cite{bax72} to construct 
the  $Q^{(1)}_{72}(v)$ which satisfies
the equations (\ref{tq1})-(\ref{tq3}) for the three cases where $m_1$
and $m_2$ are not both even. 
In sec. 6 we  consider case 4 when 
$N, m_1$ and $m_2$  are even and show that  
$Q^{(2)}_{72}(v;t)$ satisfies
(\ref{tq1}) with  additional phase factors and that in cases 4A-4D the
relations (\ref{tq2}) and (\ref{tq3}) are satisfied. 
The quasiperiodicity conditions
and the general form of the eigenvalues  of $Q_{72}^{(1)}(v),$ 
$Q^{(2)}_{72ee}(v;n\eta)$ and $Q^{(2)}_{72ee}(v;(n+1/2)\eta)$ are derived in sec.7 and
we conclude in sec. 8 with a discussion of our results.

\section{Formulation of the 8 vertex model}

The Boltzmann weights of
the 8 vertex model are given in terms of elements of a 
matrix $W_8(\alpha,\beta)_{\pm 1,\pm1}$ in a 2 dimensional  space
labeled by $\pm 1$ and an external 2 dimensional space labeled 
by $\alpha=\pm1,\beta=\pm1$. These elements are given in terms of
four quantities $a,b,c,d$ as
\begin{eqnarray}
W_8(1,1)|_{1,1}=W_8(-1,-1)|_{-1,-1}&=&a\nonumber\\
W_8(-1,-1)|_{1,1}=W_8(1,1)|_{-1,-1}&=&b\nonumber\\
W_8(-1,1)|_{1,-1}=W_8(1,-1)|_{-1,1}&=&c\nonumber\\
W_8(1,-1)|_{1,-1}=W_8(-1,1)|_{-1,1}&=&d\\
\label{bw}
\end{eqnarray}
and the transfer matrix in the $2^N\times 2^N$ ``external space''  
is written as
\begin{equation}
T_8(v)|_{\bf \alpha,\beta}={\rm Tr} W_8(\alpha_1,\beta_1)W_8(\alpha_2,\beta_2)
\cdots W_8(\alpha_N,\beta_N)
\label{t8}
\end{equation}
where the trace is in the ``internal'' $2\times 2$ space.

In the famous 1972 paper of Baxter \cite{bax72}
it is shown that any two transfer matrices commute if the 4 parameters 
for each of the two matrices are constrained by the two conditions
\begin{eqnarray}
&&\frac{a^2+b^2-c^2-d^2}{2(ab+cd)}=\Delta\\
&&\frac{cd}{ab}=\Gamma
\label{invar}
\end{eqnarray}
These two homogeneous  constraints on 4 parameters 
define a one parameter family which satisfies
\begin{equation}
[T(v),T(v')]=0
\end{equation}
The parameter $v$ is made explicit
in the 1972 paper \cite{bax72} by writing the Boltzmann weights 
in terms of the
Jacobi elliptic functions
\begin{eqnarray}
H(v)&=&2\sum_{n=1}^\infty(-1)^{n-1}q^{(n-\frac{1}{2})^2}\sin[(2n-1)\pi
v/(2K)] \label{hdef}\\
\Theta(v)&=&1+2\sum_{n=1}^{\infty}(-1)^nq^{n^2}\cos(nv\pi/K)\nonumber\\
&=&-iq^{1/4}e^{\pi i v/(2K)}H(v+iK')
\label{thetadef}
\end{eqnarray}
where $K$  and $K'$ are the standard elliptic integrals of the first kind
and 
\begin{equation}
q=e^{-\pi K'/ K}.
\end{equation}

The parametrization of \cite{bax72} is sufficient for the case 
$m_2=0$. However, to deal with
the general case of (\ref{tq4}) with $m_2\neq 0$ Baxter in 
ref. \cite{bax731} introduces
the ``modified'' theta functions 
\begin{equation}
H_m(v) = \exp\left(\frac{i\pi m_2}{8KL\eta}(v-K)^2\right)H(v)
\hspace{0.5 in}
\Theta_m(v) = \exp\left(\frac{i\pi m_2}{8KL\eta}(v-K)^2\right)\Theta(v)
\label{Hm}
\end{equation}
In terms of $\Theta_m(v)$ and $H_m(v)$ 
the Boltzmann weights are parametrized as
\begin{eqnarray}
&&a = ~~\Theta_m(-2\eta)\Theta_m(\eta-v)H_m(\eta+v) \nonumber \\
&&b = -\Theta_m(-2\eta)H_m(\eta-v)\Theta_m(\eta+v) \nonumber \\
&&c =- H_m(-2\eta)\Theta_m(\eta-v)\Theta_m(\eta+v) \nonumber \\
&&d = ~~H_m(-2\eta)H_m(\eta-v)H_m(\eta+v) 
\label{param2}
\end{eqnarray}
When $m_2=0$ the parametrization (\ref{param2}) reduces to the
parametrization of \cite{bax72}.

The factor in (\ref{Hm}) is chosen so that the modified theta
functions have the periodicity [see (eqn (11) of \cite{bax731})
\begin{equation}
H_m(v+4L\eta)=H_m(v)~~~~~~\Theta_m(v+4L\eta)=\Theta_m(v).
\label{theta4Leta}
\end{equation}
In appendix A we demonstrate that these modified theta functions are
in fact Jacobi theta functions but that their fundamental
parallelogram is no longer spanned by $2K$ and $2iK'$ (the quasiperiods
of $H(v)$ and $\Theta(v)$. Instead we find the quasiperiodicity properties
\begin{equation}
H_m(v+\omega_1) = (-1)^{r_1}(-1)^{r_1r_2}H_m(v)
\label{hm1}
\end{equation}
\begin{equation}
\Theta_m(v+\omega_1) = (-1)^{r_1r_2}\Theta_m(v)
\label{tm1}
\end{equation}
\begin{equation}
H_m(v+\omega_2) = (-1)^{b}(-1)^{ab}q'^{-1}
e^{-2\pi i(v-K)/\omega_1}H_m(v)
=(-1)^{a+b}(-1)^{ab}q'^{-1-r_2}e^{-2\pi v/\omega_1}H_m(v)
\label{hm2}
\end{equation}
\begin{equation}
\Theta_m(v+\omega_2) = (-1)^{ab}q'^{-1}
e^{-2\pi i(v-K/\omega_1)}\Theta_m(v)
=(-1)^a(-1)^{ab}q'^{-1-r_2}e^{-2\pi iv/\omega_1}\Theta_m(v)
\label{tm2}
\end{equation}
where the original (quasi)periods $2K,2iK'$ 
and the (quasi)periods $\omega_1,\omega_2$
are related by a modular transformation
\begin{equation}
\omega_1=2(r_1K+ir_2K')~~~~~~~~~\omega_2=2(bK+iaK')
\label{perioddef}
\end{equation}
\begin{equation}
ar_1-br_2=1.
\label{abdef}
\end{equation}
Here, with  $r_0$ defined as the greatest common divisor
in $2m_1$ and $m_2$,
the quantities $r_1$ and $r_2$ are given by
\begin{equation}
2m_1=r_0r_1~~~~~~~~m_2=r_0r_2,
\label{rdefs}
\end{equation}
From (\ref{abdef}) the area of the
fundamental period parallelogram
\begin{equation}
0,~\omega_1,~\omega_1+\omega_2,~ \omega_2
\label{funp}
\end{equation}
is $4KK'$.
We thus see that the modified theta functions are in fact theta
functions of nome
\begin{equation}
q'=e^{i \pi \omega_2/\omega_1}
\end{equation}
which are modular transforms
of the original theta functions $\Theta(v)$ and $H(v)$. 
We also note that note that
\begin{equation}
2L\eta=r_0\omega_1/2
\label{etaom1}
\end{equation}

\section{Formal construction of the matrices $Q_{72}(v)$}

The construction of ref. \cite{bax72} of a matrix $Q$ which satisfies
(\ref{tq1})-(\ref{tq3}) under the condition (\ref{tq4}) consists of
three steps:

\vspace{.1in}
{\bf 1. Construction of matrices $Q_R(v)$ and $Q_L(v)$}

The first step begins with an assumption  that there 
exists a matrix $Q_R(v)$
of the form
\begin{equation}
Q_R(v)|_{\alpha,\beta}=
{\rm Tr}S_R(\alpha_1,\beta_1)S_R(\alpha_2,\beta_2)\cdots S_R(\alpha_N,\beta_N)
\label{tqr0}
\end{equation}
with $S_R(\alpha,\beta)$ an $L\times L$ matrix with elements
$s_{m,n}(\alpha,\beta)$ which satisfies
\begin{equation}
T(v)Q_R(v)=[h(v+\eta)]^NQ_R(v-2\eta)+[h(v-\eta)]^NQ_R(v+2\eta)
\label{tqr1}
\end{equation}
where
\begin{equation}
h(v)=\Theta_m(0)\Theta_m(-v)H_m(v)
\label{hdef1}
\end{equation}
This matrix $Q_R(v)$ cannot be unique because if (\ref{tqr1}) is
multiplied on the left by any matrix $A$ which commutes with $T(v)$ 
then $AQ_R(v)$ also satisfies (\ref{tqr1}). In addition we note
that the matrix $e^{av}Q_R(v)$ will satisfy (\ref{tqr0}) with 
$h(v\pm\eta)^N$ 
replaced by $e^{\pm 2a\eta}h(v\pm\eta)^N$. 

Similarly we construct a matrix $Q_L(v)$
\begin{equation}
Q_L(v)|_{\alpha,\beta}=
{\rm Tr}S_L(\alpha_1,\beta_1)S_L(\alpha_2,\beta_2)\cdots S_L(\alpha_N,\beta_N)
\label{tql0}
\end{equation}
which satisfies
\begin{equation}
Q_L(v)T(v)=[h(v+\eta)]^NQ_L(v-2\eta)+[h(v-\eta)]^NQ_L(v+2\eta)
\label{tql1}
\end{equation}
This matrix is non-unique by multiplying on the right by any matrix
which commutes with $T(v)$. 

The matrices $Q_R(v)$ and $Q_L(v)$ are independently defined and can
be independently constructed by analogous procedures. However it is
also instructive to note that
the matrix $Q_L(v)$ can also be obtained obtained by taking the transpose of
(\ref{tqr1}) and using the symmetry properties of the transfer matrix
\begin{equation}
T^T(v)=(-1)^NT(-v)
\label{tmv1}
\end{equation}
\begin{equation}
T^T(v)=e^{\pi i Nm_2(v-K)/L\eta}T(2K-v) 
\label{tmv2}
\end{equation}
and the properties
\begin{equation}
h(v)=-h(-v)\label{hmv1}
\end{equation}
\begin{equation}
h(v)=e^{i\pi m_2(v-K)/(L\eta)}h(2K-v)
\label{hmv2}
\end{equation}
we find constructions for $Q_L(v)$ as
\begin{equation}
Q_L(v)=Q^{T}_R(-v)\label{qlqr1}
\end{equation}
\begin{equation}
Q_L(v)=e^{\pi im_2 vN/(2L\eta)}Q^T_R(2K-v)
\label{qlqr2}
\end{equation}
where to obtain  (\ref{qlqr1}) we have used  (\ref{tmv1})
and (\ref{hmv1}) 
where to obtain  (\ref{qlqr2}) we have used  (\ref{tmv2})
and (\ref{hmv2}) and we note that (\ref{qlqr1}) [(\ref{qlqr2})] 
may differ by right multiplication by a matrix $A$ which commutes with $T(v)$.
The matrices $Q_R(v)$ and $Q_L(v)$ will not in general satisfy
either (\ref{tq2}) or (\ref{tq3}).

\vspace{.1in}

{\bf 2. The interchange relation}

To satisfy conditions (\ref{tq2}) and (\ref{tq3}) we impose the
interchange relation
\begin{equation}
Q_L(v_1)AQ_R(v_2)=Q_L(v_2)AQ_R(v_1)
\label{commlra}
\end{equation}
where the matrix $A$ is independent of $v_1$ and $v_2,$ satisfies $A^2=1$ 
and commutes
with the transfer matrix $T(v)$. In this paper we will consider the
four choices
\begin{equation}
A=I,S,R,RS
\label{choices}
\end{equation}

These choices may be thought of as representing the arbitrariness in
the construction of $Q_R(v)$ and/or $Q_L(v).$ We will see below that for
cases 1-3 where $m_1m_2$ is not odd that (\ref{commlra}) holds for
only two of the four choices of $A$ whereas for case 4 where $m_1m_2$
is odd (\ref{commlra}) holds for all four choices (\ref{choices}).

\vspace{.1in}

{\bf 3. The nonsingularity condition}

The final requirement is that the matrices $Q_R(v)$ and $Q_L(v)$
possess  one value $v=v_0$ such that 
$Q_R(v_0)^{-1}$ and $Q_L(v_0)^{-1}$
exist. Under this nonsingularity assumption we obtain from
(\ref{tqr1}),(\ref{tql1}), and (\ref{commlra}) that the matrices
\begin{equation}
Q_{72}(v)=Q_{R}(v)Q^{-1}_R(v_0)=AQ^{-1}_L(v_0)Q_L(v)A
\label{q72}
\end{equation}
and 
\begin{equation}
AQ_{72}(v)A=AQ_{R}(v)Q^{-1}_R(v_0)A=Q^{-1}_L(v_0)Q_L(v)
\label{aq72}
\end{equation}
both satisfy the three  conditions (\ref{tq1})-(\ref{tq3}) needed for
the $TQ$ equation.

We will see below  in cases 1-3  that $Q_R(v)$ is generically
nonsingular but  for case 4 where $m_1m_2$ is odd
and (\ref{commlra}) holds for all four choices (\ref{choices}) that
$Q_R(v)$ is singular for all $v$. In cases 1-3 where the  interchange
relation (\ref{commlra}) holds for two and only two matrices $A_1$ and $A_2$
we obtain from (\ref{aq72})
\begin{equation}
A_1Q_{72}(v)A_1=A_2Q_{72}(v)A_2
\end{equation}
or
\begin{equation}
A_2A_1Q_{72}(v)A_1A_2= Q_{72}(v)
\end{equation}
If $A_1,A_2$ commute then $Q_{72}(v)$ commutes with $A_1A_2$

\section{The matrices $Q_R(v)$ and $Q_L(v)$}
In appendix C of ref. \cite{bax72} it is shown that for the  existence
of the matrix $Q_R(v)$ of the form (\ref{tqr0}) which satisfies
  (\ref{tqr1}) it is necessary that the matrix elements $s^R_{m,n}$ of
$S_R$ satisfy
\begin{eqnarray}
&&(ap_n-bp_m)s^R_{m,n}(+,\beta)+(d-cp_mp_n)s^R_{m,n}(-,\beta)=0\nonumber\\
&&(c-dp_mp_n)s^R_{m,n}(+\beta)+(bp_n-ap_m)s^R_{m,n}(-,\beta)=0
\label{tqr2}
\end{eqnarray}
This set of homogeneous linear equations will have a 
nontrivial solution provided
\begin{equation}
(a^2+b^2-c^2-d^2)p_mp_n=ab(p_m^2+p_n^2)-cd(1+p_m^2p_n^2)
\label{tqr3}
\end{equation}
This can only happen for certain values of $m$ and $n$. For all other
values we have
\begin{equation}
s^R_{m,n}(\alpha,\beta)=0
\label{tqr4}
\end{equation}
Using the parameterizations (\ref{param2})
we have 
\begin{equation}
\frac{a^2+b^2-c^2-d^2}{ab}=2{\rm cn}(2\eta){\rm dn}(2\eta),~~~
cd/ab=k{\rm sn}^2(2\eta)\label{ident}
\end{equation}
where ${\rm sn}(v),{\rm cn}(v), {\rm dn}(v)$ 
are the conventional doubly periodic
functions with periods $2K$ and $2iK'$ and ${\rm sn}(v)$ is given in terms
of theta functions as
\begin{equation}
k^{1/2}{\rm sn}(v)=H(v)/\Theta(v)
\end{equation}
Thus (\ref{tqr3}) becomes
\begin{equation}
2{\rm cn}(2\eta){\rm dn}(2\eta)p_mp_n
=p^2_m+p^2_n-k{\rm sn}^2(2\eta)(1+p_m^2p_n^2)  
\label{ntqr3}
\end{equation}
 and it is shown in ref.\cite{bax72} 
that if $p_m$ is written as
\begin{equation}
p_m=k^{1/2}sn(u)
\label{pm}
\end{equation}
it follows from (\ref{ntqr3}) that
\begin{equation}
p_n=k^{1/2}sn(u\pm 2\eta)
\label{pn}
\end{equation}
In order for the nonsingularity condition for $Q_R(v)$ to hold we need
additional nonvanishing elements $s^R_{m,n}$. We consider two possible
choices.

\subsection{The matrices $Q^{(1)}_R(v)$ and $Q^{(1)}_L(v)$}
The first choice is to require
\begin{equation}
s_{1,1}(\alpha,\beta)\neq 0,~~~s_{L,L}(\alpha,\beta)\neq 0
\label{choice1}
\end{equation}
Then equ. (\ref{ntqr3}) has to be satisfied for $n=m$. Then
\begin{equation}
sn(u) = sn(u \pm 2\eta)
\end{equation}
This fixes the parameter $u$ to become $u=K \pm \eta$ and leads to the restriction to discrete $\eta$:
\begin{equation}
2L\eta=2m_1K+im_2K'
\label{rou0}
\end{equation}
One obtains from (\ref{pm}) and (\ref{pn}) that
\begin{equation}
p_n=k^{1/2}sn(K+(2n-1)\eta)
\label{pnn}
\end{equation}
We indicate the choice (\ref{choice1}) by writing 
\begin{equation}
S_R(\alpha,\beta)\rightarrow S_R^{(1)}(\alpha,\beta),~~~Q_R(v)\rightarrow
Q_R^{(1)}(v)
\end{equation}
and from \cite{bax72} we find
\begin{equation}
\begin{array}{lclcl} 
 S^{(1)}_R(+,\beta)_{k,k+1}(v) & = &   H_m(v+K-2k\eta)\tau_{\beta,-k}&~~&1\leq
 k\leq L-1\\
S^{(1)}_R(+,\beta)_{k+1,k}(v) & = &  H_m(v+K+2k\eta)\tau_{\beta,~k}&~~&1\leq
 k\leq L-1\\
S^{(1)}_R(+,\beta)_{1,1}(v)   & = &  H_m(v+K)\tau_{\beta,0}&~~&\\
S^{(1)}_R(+,\beta)_{L,L}(v)   & = &  H_m(v+K+2L\eta)\tau_{\beta,L}&~~&\\
S^{(1)}_R(-,\beta)_{k,k+1}(v) & = &  \Theta_m(v+K-2k\eta)\tau_{\beta,-k}&~~&1\leq
 k\leq L-1\\
S^{(1)}_R(-,\beta)_{k+1,k}(v) & = &  \Theta_m(v+K+2k\eta)\tau_{\beta,~k}&~~&1\leq
 k\leq L-1\\
S^{(1)}_R(-,\beta)_{1,1}(v)   & = & \Theta_m(v+K)\tau_{\beta,0}&~~&\\
S^{(1)}_R(-,\beta)_{L,L}(v)   & = &  \Theta_m(v+K+2L\eta)\tau_{\beta,L}&~~&\\
\end{array}
\label{SRm}
\end{equation}
With this choice the argument of appendix C of ref. \cite{bax72} shows
that the equation (\ref{tqr1}) holds with $h(v)$ given by (\ref{hdef1}).

We choose to construct the matrix $Q_L(v)$ from (\ref{SRm})
by use of (\ref{qlqr1}) and (\ref{help1}) as
\begin{eqnarray}
\begin{array}{lclcl}
 S^{(1)}_L(\alpha,+)_{k,k+1}(v) 
& = &  H_m(v+K+2k\eta)\tau'_{\alpha,-k}&~~&1\leq
 k\leq L-1\\
 S^{(1)}_L(\alpha,+)_{k+1,k}(v) 
& = &  H_m(v+K-2k\eta)\tau'_{\alpha,~k}&~~&1\leq
 k\leq L-1\\
 S^{(1)}_L(\alpha,+)_{1,1}(v)   
& = &  H_m(v+K)\tau'_{\alpha,0}&~~&\\
 S^{(1)}_L(\alpha,+)_{L,L}(v)   
& = &  H_m(v+K-2L\eta)\tau'_{\alpha,L}&~~&\\
 S^{(1)}_L(\alpha,-)_{k,k+1}(v) 
& = & \Theta_m(v+K+2k\eta)\tau'_{\alpha,-k}&~~&1\leq
 k\leq L-1\\
 S^{(1)}_L(\alpha,-)_{k+1,k}(v) 
& = & \Theta_m(v+K-2k\eta)\tau'_{\alpha,~k}&~~&1\leq
 k\leq L-1\\
 S^{(1)}_L(\alpha,-)_{1,1}(v)   
& = & \Theta_m(v+K)\tau'_{\alpha,0}&~~&\\
 S^{(1)}_L(\alpha,-)_{L,L}(v)   
& = & \Theta_m(v+K-2L\eta)\tau'_{\alpha,L}&~~&\\
\end{array}
\label{SL}
\end{eqnarray}
In order to cover the case $m_2 \neq 0$ we have to define $S_R$ and $S_L$ in terms
of modified theta functions. That this is allowed is immediately obvious from the
simple observation that 
\begin{equation}
k^{1/2}{\rm sn}(v)=H(v)/\Theta(v)=H_m(v)/\Theta_m(v)
\end{equation}

\subsection{The matrices $Q^{(2)}_R(v;t)$ and $Q^{(2)}_L(v;t)$ for $N$ even}

The second choice which exists for $m_1$ and $m_2$ even 
was recently found in ref. \cite{klaus} with
\begin{equation}
s^R_{1,L}(\alpha,\beta)\neq 0,~~~s^R_{L,1}(\alpha,\beta)\neq 0
\label{choice2}
\end{equation}
This choice will always give a vanishing matrix $Q^{(2)}_R$
when used in (\ref{tqr0}) when $N$ is odd. Consequently
whenever we consider $Q^{(2)}_R(v;t)$ we will always assume that $N$ is
even.

To obtain this case
we need to have (\ref{ntqr3}) hold for  
$m=1,n=L$ and $m=L,n=1$ which
because of the symmetry in (\ref{ntqr3}) in $m$ and $n$ gives the
single equation 
\begin{eqnarray}
&&{\rm sn}^2(v_r+2\eta)+{\rm sn}^2(v_r+2L\eta)-{\rm sn}^2
2\eta(1+k^2{\rm sn}^2(v_r+2\eta){\rm sn}^2(v_r+2L\eta))\nonumber\\
&&-2{\rm
  sn}(v_r+2\eta){\rm sn}(v_r+2L\eta){\rm cn} 2\eta{\rm dn}2\eta=0
\end{eqnarray}
This equation will hold if $p_n$ is given by (\ref{pn})  
with $p_1=p_{L+1}$ and thus
\begin{equation}
{\rm sn}(v_r+2\eta)={\rm sn}(v_r+2(L+1)\eta)
\label{tq28}
\end{equation}
which, using the periodicity properties ${\rm sn}(v+2K)=-{\rm sn}v$ and
${\rm sn}(v+2iK')={\rm sn}v,$ is satisfied for all $v$ if
\begin{equation}
2L\eta=4{\tilde m}_1K+2i{\tilde m}_2K'
\label{tq29}
\end{equation} 
which is the root of unity condition (\ref{tq4}) with $m_1=2{\tilde
  m}_1,~m_2=2{\tilde m}_2$. In other words we are restricted to $m_1$
  and $m_2$ even in the root of unity condition (\ref{tq4}).

We will follow the notation of ref. \cite{klaus} by setting
\begin{equation}
v_r=t-\eta
\end{equation}
Then, indicating the choice (\ref{choice2})  by writing
\begin{equation}
S_R(\alpha,\beta)\rightarrow S_R^{(2)}(\alpha,\beta),~~~Q_R(v)\rightarrow
Q_R^{(2)}(v)
\end{equation}
we have
\begin{equation}
p_n=k^{1/2}{\rm sn}[t+(2n-1)\eta]=H_m(t+(2n-1)\eta)/\Theta_m(t+(2n-1)\eta)
\label{pn2}
\end{equation}
and following \cite{klaus} we find
\begin{equation}
\begin{array}{lcl} 
S^{(2)}_R(+,\beta)_{k,k+1}(v) & = &  - H_m(v-t-2k\eta)\tau_{\beta,-k}\\
S^{(2)}_R(+,\beta)_{k+1,k}(v) & = & ~ H_m(v+t+2k\eta)\tau_{\beta,~k}\\
S^{(2)}_R(-,\beta)_{k,k+1}(v) & = &  ~\Theta_m(v-t-2k\eta)\tau_{\beta,-k}\\
S^{(2)}_R(-,\beta)_{k+1,k}(v) & = &  ~\Theta_m(v+t+2k\eta)\tau_{\beta,~k}\\
S^{(2)}_R(+,\beta)_{1,L}(v)&=&H_m(v+t+2L\eta)\tau_{\beta,~L}\\
S^{(2)}_R(+,\beta)_{L,1}(v)&=&-H_m(v-t-2L\eta)\tau_{\beta,-L}\\
S^{(2)}_R(-,\beta)_{1,L}(v) & = &~\Theta_m(v+t+2L\eta)\tau_{\beta,L}\\
S^{(2)}_R(-,\beta)_{L,1}(v) & = &  ~\Theta_m(v-t-2L\eta)\tau_{\beta,-L}\\
\end{array}
\label{SRm2}
\end{equation}

With the choice (\ref{SRm2}) for $S^{(2)}_R$ we may follow the
procedure of \cite{bax72} to obtain the slight generalization of (\ref{tq1})
\begin{equation}
T(v)Q^{(2)}_R(v;t)=\omega^{-N}[h(v+\eta)]^NQ^{(2)}_R(v-2\eta;t)
+\omega^{-N}[h(v-\eta)]^NQ^{(2)}_R(v+2\eta;t)
\label{2tq1}
\end{equation}
where
\begin{equation}
\omega={\rm exp}\left(\frac{i\pi m_2}{2L}\right)
\label{2tq2}
\end{equation}
The details of this computation which show the origin of the phase
factor $\omega$ are given in appendix C.

The companion matrix $Q^{(2)}_L(v;t)$ must satisfy
\begin{equation}
Q_L^{(2)}(v,t)T(v)=\omega^{-N}[h(v+\eta)]^NQ^{(2)}_L(v-2\eta;t)
+\omega^{-}[h(v-\eta)]^NQ^{(2)}_L(v+2\eta;t)
\label{t2tq3}
\end{equation}
We find it convenient to use (\ref{tmv1}) and (\ref{hmv1}) to construct
$Q^{(2)}_L(v;t)$ in terms of $Q^{(2)}_R(v;t)$ as
\begin{equation}
Q^{(2)}_L(v;t)=-Q_R^{(2)T}(2K-v;t)S
\end{equation}
where the factor of $-S$ which  is inserted for convenience uses the
non uniqueness of $Q_L(v)$ under multiplication on the right by any
matrix which commutes with $T(v).$
Thus we find
\begin{equation}
\begin{array}{lcl} 
 S^{(2)}_L(\alpha,+)_{k,k+1}(v) & = &   H_m(v+t+2k\eta)\tau'_{\alpha,-k}\\
S^{(2)}_L(\alpha,+)_{k+1,k}(v) & = &  -H_m(v-t-2k\eta)\tau'_{\alpha,~k}\\
S^{(2)}_L(\alpha,-)_{k,k+1}(v) & = &  \Theta_m(v+t+2k\eta)\tau'_{\alpha,-k}\\
S^{(2)}_L(\alpha,-)_{k+1,k}(v) & = &  \Theta_m(v-t-2k\eta)\tau'_{\alpha,~k}\\
 S^{(2)}_L(\alpha,+)_{1,L}(v) & = &  -H_m(v-t-2L\eta)
\tau'_{\alpha,L}\\
S^{(2)}_L(\alpha,+)_{L,1}(v) & = & H_m(v+t+2L\eta)\tau'_{\alpha,-L}\\
S^{(2)}_L(\alpha,-)_{1,L}(v) & = & \Theta_m(v-t-2L\eta)\tau'_{\alpha,L}\\
S^{(2)}_L(\alpha,-)_{L,1}(v) & = &  \Theta_m(v+t+2L\eta)
\tau'_{\alpha,-L}\end{array}\\
\label{sl2}
\end{equation}

\section{The matrices $Q_{72}^{(1)}(v)$ for $m_1$ and $m_2$ not both even}
 
The construction of the matrices $Q^{(1)}_R(v)$ and $Q^{(1)}_L(v)$ 
given in the previous section is valid for all integer $m_1$ and $m_2$
in the root of unity condition (\ref{tq4}). However, the validity and
the choice of the matrix $A$ in the interchange relation (\ref{commlra})
and the nonsingularity condition are different for the different
parities of $m_1$ and $m_2$.   

\subsection{The interchange relations}

The computation of the interchange relations (\ref{commlra}) are similar
for all four choices of the matrix $A$ but each case differs in
detail. Therefore we will treat the four cases in separately. 
The results are summarized in 4.1.5. 

\subsubsection{The case $A=I$}

We consider first the interchange  relation (\ref{commlra}) with $A=1$
and write
\begin{equation}
Q_L^{(1)}(v')Q_R^{(1)}(v)|_{\alpha,\beta}
={\rm Tr}W^{(1)}(\alpha_1,\beta_1|v',v)\cdots W^{(1)}(\alpha_N,\beta_N|v',v)
\label{proof1}
\end{equation} 
where $W^{(1)}(\alpha,\beta|v',v)$ are $L^2\times L^2$ matrices with elements
\begin{equation}
W^{(1)}(\alpha,\beta|v',v)_{k,k';l,l'}=\sum_{\gamma=\pm}
S^{(1)}_L(\alpha,\gamma|v')_{k,l}
S^{(1)}_R(\gamma,\beta|v)_{k',l'}
\label{proof2}
\end{equation}
Thus the  interchange  relation (\ref{commlra}) with $A=I$ 
will follow if we can show that 
there exists an $L^2\times L^2$  diagonal matrix $Y$ with elements
\begin{equation}
y^{(1)}_{k,k';l,l'}=y^{(1)}_{k,k'}\delta_{k,l}\delta_{k',k'}
\end{equation}
such that
\begin{equation}
W^{(1)}(\alpha,\beta|v',v)=Y^{(1)}W^{(1)}(\alpha,\beta|v,v')Y^{(1)-1}
\end{equation}

To examine the possibility of the existence of such a diagonal similarity
transformation we need to explicitly compute $W(\alpha,\beta|v',v)$ from
(\ref{proof2}). To do this we use the identity
\begin{equation}
\Theta_m(v')\Theta_m(v)+H_m(v')H_m(v)=f_{+}(v+v')g_{+}(v'-v)
\label{proof3}
\end{equation}
with
\begin{equation}
f_{+}(z) = -\frac{2q^{1/4}}{H(K)\Theta(K)}\exp\left(\frac{i\pi m_2}{8KL\eta}(K'^2+2iKK'-2Kz)\right)
H_m((iK'+z)/2)H_m((iK'-z)/2)
\label{proof4}
\end{equation}
\begin{equation}
g_{+}(z) =H_{m}((iK'+z)/2+K)H_{m}((iK'-z)/2+K)
\label{proof5}
\end{equation}
where we note the following properties
\begin{eqnarray}
&&f_{+}(-z)=e^{i\pi m_2 z/(2L\eta)}f_{+}(z)\label{fpprop}\\
&&g_{+}(-z)=g_{+}(z)\label{gppropm}\\
&&g_{+}(z+4L\eta)=(-1)^{m_1m_2}g_{+}(z)\label{proof14}
\end{eqnarray}
and for $m_1$ and $m_2$ both even
\begin{equation}
g_{+}(v+2L\eta)=(-1)^{m_1m_2/4}g_{+}(z)
\label{newgpprop}
\end{equation}
The properties (\ref{fpprop}) and (\ref{gppropm}) are obvious from the
definitions (\ref{proof4}) and (\ref{proof5}).  
Property (\ref{newgpprop}) follows from (\ref{proof14})
with $m_1\rightarrow m_1/2$ and $m_2\rightarrow m_2/2$. 
The relation (\ref{proof14}) follows from (\ref{Hminus}),(\ref{more1}),(\ref{more2})
and (\ref{Thsh}).
Using (\ref{proof3})  we find explicitly
\begin{eqnarray}
&&W^{(1)}(\alpha,\beta|v',v)_{k.k',l.l'}  \nonumber \\
&&=\delta_{k+1,l}\delta_{k'+1,l'}\tau'_{\alpha,-k}\tau_{\beta,-k'}
f_{+}(v'+v+2K+2(k-k')\eta)g_{+}(v'-v+2(k+k')\eta) \nonumber \\
&&+\delta_{k+1,l}\delta_{k',l'+1}\tau'_{\alpha,-k}\tau_{\beta,l'}
f_{+}(v'+v+2K+2(k+l')\eta)g_{+}(v'-v+2(k-l')\eta) \nonumber \\
&&+\delta_{k+1,l}\delta_{k',1'}\delta_{l',1}\tau'_{\alpha,-k}\tau_{\beta,0}
f_{+}(v'+v+2K+2k\eta)g_{+}(v'-v+2k\eta) \nonumber \\
&&+\delta_{k+1,l}\delta_{k',L}\delta_{l',L}\tau'_{\alpha,-k}\tau_{\beta,L}
f_{+}(v'+v+2K+2(k+L)\eta)g_{+}(v'-v+2(k-L)\eta) \nonumber \\
&&+\delta_{k,l+1}\delta_{k'+1,l'}\tau'_{\alpha,l}\tau_{\beta,-k'}
f_{+}(v'+v+2K-2(l+k')\eta)g_{+}(v'-v-2(l-k')\eta) \nonumber \\
&&+\delta_{k,l+1}\delta_{k',l'+1}\tau'_{\alpha,l}\tau_{\beta,l'}
f_{+}(v'+v+2K-2(l-l')\eta)g_{+}(v'-v-2(l+l')\eta) \nonumber \\
&&+\delta_{k,l+1}\delta_{k',1'}\delta_{l',1}\tau'_{\alpha,l}\tau_{\beta,0}
f_{+}(v'+v+2K-2l\eta)g_{+}(v'-v-2l\eta) \nonumber \\
&&+\delta_{k,l+1}\delta_{k',L}\delta_{l',L}\tau'_{\alpha,l}\tau_{\beta,L}
f_{+}(v'+v+2K-2(l-L)\eta)g_{+}(v'-v-2(l+L)\eta) \nonumber \\
&&+\delta_{k,1}\delta_{l,1}\delta_{k'+1,l'}\tau'_{\alpha,0}\tau_{\beta,-k'}
f_{+}(v'+v+2K-2k'\eta)g_{+}(v'-v+2k'\eta)\nonumber \\
&&+\delta_{k,1}\delta_{l,1}\delta_{k',l'+1}\tau'_{\alpha,0}\tau_{\beta,l'}
f_{+}(v'+v+2K+2l'\eta)g_{+}(v'-v-2l'\eta)\nonumber \\
&&+\delta_{k,L}\delta_{l,L}\delta_{k'+1,l'}\tau'_{\alpha,L}\tau_{\beta,-k'}
f_{+}(v'+v+2K-2(L+k')\eta)g_{+}(v'-v-2(L-k')\eta)\nonumber \\
&&+\delta_{k,L}\delta_{l,L}\delta_{k',l'+1}\tau'_{\alpha,L}\tau_{\beta,l'}
f_{+}(v'+v+2K-2(L-l')\eta)g_{+}(v'-v-2(L+l')\eta)\nonumber \\
&&+\delta_{k,1}\delta_{l,1}\delta_{k',1'}\delta_{l',1}
\tau'_{\alpha,0}\tau_{\beta,0}
f_{+}(v'+v+2K)g_{+}(v'-v)\nonumber \\
&&+\delta_{k,1}\delta_{l,1}\delta_{k',L}\delta_{l',L}
\tau'_{\alpha,0}\tau_{\beta,L}
f_{+}(v'+v+2K+2L\eta)g_{+}(v'-v-2L\eta)\nonumber \\
&&+\delta_{k,L}\delta_{l,L}\delta_{k',1'}\delta_{l',1}
\tau'_{\alpha,L}\tau_{\beta,0}
f_{+}(v'+v+2K-2L\eta)g_{+}(v'-v-2L\eta)\nonumber \\
&&+\delta_{k,L}\delta_{l,L}\delta_{k',L}\delta_{l',L}
\tau'_{\alpha,L}\tau_{\beta,L}
f_{+}(v'+v+2K)g_{+}(v'-v-4L\eta)
\label{proof7}
\end{eqnarray}

A necessary condition for the existence of a diagonal similarity transformation
is that the diagonal elements $W^{(1)}(\alpha,\beta|v',v)_{k,k';k,k'}$ 
and $W^{(1)}(\alpha,\beta|v,v')_{k.k':k,k'}$ be equal. 
From the last four terms in (\ref{proof7})  we find that
these diagonal elements are
\begin{equation}
W^{(1)}(\alpha,\beta|v',v)_{11,11} = f_{+}(v'+v+2K)g_{+}(v'-v)
\tau'_{\alpha,0}\tau_{\beta,0}
\label{proof8}
\end{equation}
\begin{equation}
W^{(1)}(\alpha,\beta|v',v)_{1,L;1,L} 
= f_{+}(v'+v+2K+2L\eta)g_{+}(v'-v-2L\eta)\tau'_{\alpha,0}
\tau_{\beta,L}
\label{proof9}
\end{equation}
\begin{equation}
W^{(1)}(\alpha,\beta|v',v)_{L,1;L,1}=f_{+}(v'+v+2K-2L\eta)g_{+}(v'-v-2L\eta)
\tau'_{\alpha,L}\tau_{\beta,0}
\label{proof10}
\end{equation}
\begin{equation}
W(\alpha,\beta|v',v)_{L,L;L,L}=f_{+}(v'+v+2K)g_{+}(v'-v-4L\eta)
\tau'_{\alpha,L}\tau_{\beta,L}
\label{proof11}
\end{equation}

It follows from (\ref{proof8}) and (\ref{proof11})
by use of (\ref{proof14}) that
\begin{eqnarray}
&&W^{(1)}(\alpha,\beta|v',v)_{1,1;1,1}=W^{(1)}(\alpha,\beta|v,v')_{1.1;1,1}
\label{proof12}\\
&&W^{(1)}(\alpha,\beta|v',v)_{L,L;L,L}=W^{(1)}(\alpha,\beta|v,v')_{L,L;L,L}
\label{proof13}
\end{eqnarray}

To examine the elements $W^{(1)}(\alpha,\beta|v',v)_{1,L;1,L}$ and 
$W^{(1)}(\alpha,\beta|v',v)_{L,1;L,1}$
we use the identity (\ref{proof14}) in
(\ref{proof9}) and (\ref{proof10}) we find
\begin{eqnarray}
&&W^{(1)}(\alpha,\beta|v',v)_{1,L;1,L}
=(-1)^{m_1m_2}W^{(1)}(\alpha,\beta|v,v')_{1,L;1,L}
\label{proof18}\\
&&W^{(1)}(\alpha,\beta|v',v)_{L,1;L,1}
=(-1)^{m_1m_2}W^{(1)}(\alpha,\beta|v,v')_{L,1;L,1}
\label{proof19}
\end{eqnarray}
 
We thus conclude that the interchange relation (\ref{commlra}) is not
satisfied if $m_1$ and $m_2$ are both odd.
However, when at least one of the integers $m_1,m_2$ is even we do
have the necessary equality 
and the remainder of the proof of
the existence of the diagonal matrix $Y$ as given in appendix C of
ref. \cite{bax72} holds.  We thus conclude that in the case that 
when $m_1m_2$ is even that (\ref{commlra}) holds.
 
\subsubsection{The case $A=S$}

The proof of the  interchange relation (\ref{commlra}) 
with $A=S$ is similar to the proof with $A=I$

We first write
\begin{equation}
Q_L^{(1)}(v')SQ_R^{(1)}(v)|_{\alpha,\beta}
={\rm Tr}W^{(1)S}(\alpha_1,\beta_1|v',v)\cdots W^{(1)S}(\alpha_N,\beta_N|v',v)
\label{proofs1}
\end{equation} 
where $W^{(1)S}(\alpha,\beta|v',v)$ are $L^2\times L^2$ matrices with elements
\begin{equation}
W^{(1)S}(\alpha,\beta|v',v)_{k,k';l,l'}
=\sum_{\gamma=\pm}\gamma S_L(\alpha,\gamma|v')_{k,l} 
S_R(\gamma,\beta|v)_{k',l'}
\label{proofs2}
\end{equation}
Thus (\ref{commlra}) will follow if we can show that 
there exists an $L^2\times L^2$  diagonal matrix $Y^S$ with elements
\begin{equation}
y^{(1)S}_{k,k';l,l'}=y^{(1)S}_{k,k'}\delta_{k,l}\delta_{k',l'}
\label{proofs3}
\end{equation}
such that
\begin{equation}
W^{(1)S}(\alpha,\beta|v',v)=Y^{(1)S}W^S(\alpha,\beta|v,v')Y^{(1)S-1}
\label{proofs4}
\end{equation}
To explicitly compute the matrix $W^{(1)S}(\alpha,\beta|v',v)$ 
we use the identity which follows immediately from the identity
(\ref{proof3}) and (\ref{Hminus})
by sending $v'\rightarrow -v'$ 
\begin{equation}
H_m(u)H_m(v)-\Theta_m(u)\Theta_m(v) = f_{-}(u+v)g_{-}(u-v)
\label{proofs5}
\end{equation}
where
\begin{equation}
f_{-}(z) = \frac{2q^{1/4}}{H(K)\Theta(K)}\exp\left(\frac{i\pi m_2}{8KL\eta}(K'^2+2iKK'-2Kz)\right)
H_m((iK'+z)/2+K)H_m((iK'-z)/2+K)
\label{proofs6}
\end{equation}
\begin{equation}
g_{-}(z) = H_{m}((iK'+z)/2)H_{m}((iK'-z)/2)
\label{proofs7}
\end{equation}
which have the properties that
\begin{eqnarray}
&&f_{-}(-z)=e^{i \pi m_2 z/(2L\eta)}f_{-}(z)\label{proofs8}\\
&&g_{-}(-z)=g_{-}(z)\label{gmpropm}\\
&&g_{-}(z+4L\eta)=(-1)^{m_1m_2}(-1)^{m_2}g_{-}(z)\label{proofs17}
\end{eqnarray}
and for $m_1$ and $m_2$ both even
\begin{equation}
g_{-}(z+2L\eta)=(-1)^{m_1m_2/4}(-1)^{m_2/2}g_{-}(z)\label{newgmprop}
\end{equation}
The property (\ref{newgmprop}) follows from (\ref{proofs17})
with $m_1\rightarrow m_1/2$ and $m_2\rightarrow m_2/2$.
The proof of (\ref{proofs17}) follows from (\ref{Hminus}),(\ref{more1}),(\ref{more2})
and (\ref{Thsh}).
The properties (\ref{proofs8}) and (\ref{gmpropm}) are obvious from the
definitions (\ref{proofs6}) and (\ref{proofs7})
\begin{eqnarray}
&&W^{(1)S}(\alpha,\beta|v',v)_{k,k';l,l'}=\nonumber\\
&&+\delta_{k+1,l}\delta_{k'+1,l'}\tau'_{\alpha,-k}\tau_{\beta,-k'}
f_{-}(v'+v+2K+2(k-k')\eta)g_{-}(v'-v+2(k+k')\eta) \nonumber \\
&&+\delta_{k+1,l}\delta_{k',l'+1}\tau'_{\alpha,-k}\tau_{\beta,l'}
f_{-}(v'+v+2K+2(k+l')\eta)g_{-}(v'-v+2(k-l')\eta) \nonumber \\
&&+\delta_{k+1,l}\delta_{k',1'}\delta_{l',1}\tau'_{\alpha,-k}\tau_{\beta,0}
f_{-}(v'+v+2K+2k\eta)g_{-}(v'-v+2k\eta) \nonumber \\
&&+\delta_{k+1,l}\delta_{k',L}\delta_{l',L}\tau'_{\alpha,-k}\tau_{\beta,L}
f_{-}(v'+v+2K+2(k+L)\eta)g_{-}(v'-v+2(k-L)\eta) \nonumber \\
&&+\delta_{k,l+1}\delta_{k'+1,l'}\tau'_{\alpha,l}\tau_{\beta,-k'}
f_{-}(v'+v+2K-2(l+k')\eta)g_{-}(v'-v-2(l-k')\eta) \nonumber \\
&&+\delta_{k,l+1}\delta_{k',l'+1}\tau'_{\alpha,l}\tau_{\beta,l'}
f_{-}(v'+v+2K-2(l-l')\eta)g_{-}(v'-v-2(l+l')\eta) \nonumber \\
&&+\delta_{k,l+1}\delta_{k',1'}\delta_{l',1}\tau'_{\alpha,l}\tau_{\beta,0}
f_{-}(v'+v+2K-2l\eta)g_{-}(v'-v-2l\eta) \nonumber \\
&&+\delta_{k,l+1}\delta_{k',L}\delta_{l',L}\tau'_{\alpha,l}\tau_{\beta,L}
f_{-}(v'+v+2K-2(l-L)\eta)g_{-}(v'-v-2(l+L)\eta) \nonumber \\
&&+\delta_{k,1}\delta_{l,1}\delta_{k'+1,l'}\tau'_{\alpha,0}\tau_{\beta,-k'}
f_{-}(v'+v+2K-2k'\eta)g_{-}(v'-v+2k'\eta)\nonumber \\
&&+\delta_{k,1}\delta_{l,1}\delta_{k',l'+1}\tau'_{\alpha,0}\tau_{\beta,l'}
f_{-}(v'+v+2K+2l'\eta)g_{-}(v'-v-2l'\eta)\nonumber \\
&&+\delta_{k,L}\delta_{l,L}\delta_{k'+1,l'}\tau'_{\alpha,L}\tau_{\beta,-k'}
f_{-}(v'+v+2K-2(L+k')\eta)g_{-}(v'-v-2(L-k')\eta)\nonumber \\
&&+\delta_{k,L}\delta_{l,L}\delta_{k',l'+1}\tau'_{\alpha,L}\tau_{\beta,l'}
f_{-}(v'+v+2K-2(L-l')\eta)g_{-}(v'-v-2(L+l')\eta)\nonumber \\
&&+\delta_{k,1}\delta_{l,1}\delta_{k',1'}\delta_{l',1}
\tau'_{\alpha,0}\tau_{\beta,0}
f_{-}(v'+v+2K)g_{-}(v'-v)\nonumber \\
&&+\delta_{k,1}\delta_{l,1}\delta_{k',L}\delta_{l',L}
\tau'_{\alpha,0}\tau_{\beta,L}
f_{-}(v'+v+2K+2L\eta)g_{-}(v'-v-2L\eta)\nonumber \\
&&+\delta_{k,L}\delta_{l,L}\delta_{k',1'}\delta_{l',1}
\tau'_{\alpha,L}\tau_{\beta,0}
f_{-}(v'+v+2K-2L\eta)g_{-}(v'-v-2L\eta)\nonumber \\
&&+\delta_{k,L}\delta_{l,L}\delta_{k',L}\delta_{l',L}
\tau'_{\alpha,L}\tau_{\beta,L}
f_{-}(v'+v+2K)g_{-}(v'-v-4L\eta)\nonumber 
\label{proofs10}
\end{eqnarray}

In order for the diagonal matrix $Y^{(1)S}$ to exist it is necessary
  that the diagonal elements of $W^{(1)S}(\alpha,\beta|v',v)_{k,k';k,k'}$ be
  symmetric under the interchange of $v'$ and $v$.
From the last four terms in (\ref{proofs10}) these diagonal elements are
\begin{equation}
W^{(1)S}(\alpha,\beta|v',v)_{1,1;1,1}= f_{-}(v'+v+2K)g_{-}(v'-v)
\tau'_{\alpha,0}\tau_{\beta,0}
\label{proofs11}
\end{equation}
\begin{equation}
W^{(1)S}(\alpha,\beta|v',v)_{1,L;1,L}= f_{-}(v'+v+2K+2L\eta)
g_{-}(v'-v-2L\eta)\tau'_{\alpha,0}\tau_{\beta,L}
\label{proofs12}
\end{equation}
\begin{equation}
W^{(1)S}(\alpha,\beta|v',v)_{L,1;L,1}=
f_{-}(v'+v+2K-2L\eta)g_{-}(v'-v-2L\eta)\tau'_{\alpha,L}\tau_{\beta,0}
\label{proofs13}
\end{equation}
\begin{equation}
W^{(1)S}(\alpha,\beta|v',v)_{L,L;L,L}=f_{-}(v'+v+2K)g_{-}(v'-v-4L\eta)
\tau'_{\alpha,L}\tau_{\beta,L}
\label{proofs14}
\end{equation}

The equalities
\begin{eqnarray}
&&W^{(1)S}(\alpha,\beta|v',v)_{1,1;1,1}=W^{(1)S}(\alpha,\beta|v,v')_{1,1;1,1}\label{proofs15}\\
&&W^{(1)S}(\alpha,\beta|v',v)_{L,L;L,L}=W^{(1)S}(\alpha,\beta|v,v')_{L,L;L,L}
\label{proofs16}
\end{eqnarray}
follow from (\ref{proofs17}).

To study $W^{(1)S}(\alpha,\beta|v',v)_{1,L;1L}$ 
and $W^{(1)S}(\alpha,\beta|v',v)_{L,1;L,1}$
we use the identity (\ref{proofs17})
in (\ref{proofs12}) and (\ref{proofs13})
to obtain 
\begin{eqnarray}
&&W^{(1)S}(\alpha,\beta|v',v)_{1,L;1,L}
=(-1)^{m_1m_2}(-1)^{m_2}W^{(1)S}(\alpha,\beta|v,v')_{1,L;1,L}
\label{proofs21}\\
&&W^{(1)S}(\alpha,\beta|v',v)_{L,1;L,1}
=(-1)^{m_1m_2}(-1)^{m_2}W^{(1)S}(\alpha,\beta|v,v')_{L,1;L,1}
\label{proofs22}
\end{eqnarray}
From (\ref{proofs21}) and (\ref{proofs22}) we conclude that
in order for (\ref{commlra}) to hold with $A=S$  both $m_1$ and $m_2$ must be
odd or $m_2$ must be even. With this restriction the method of appendix C
of ref. \cite{bax72} demonstrates the existence of the similarity
transformation $Y$ and thus (\ref{commlra})with $A=S$ is proven.

\subsubsection{The case $A=R$}

For the case $A=R$ we consider
\begin{equation}
Q_L^{(1)}(v')RQ_R^{(1)}(v)|_{\alpha,\beta}
={\rm Tr}W^{(1)R}(\alpha_1,\beta_1|v',v)\cdots W^{(1)R}(\alpha_N,\beta_N|v',v)
\label{proof1r}
\end{equation} 
where $W^{(1)R}(\alpha,\beta|v',v)$ are $L^2\times L^2$ matrices with elements
\begin{eqnarray}
&&W^{(1)R}(\alpha,\beta|v',v)_{k,k';l,l'}=\sum_{\gamma=\pm}
S^{(1)}_L(\alpha,\gamma|v')_{k,l}
S^{(1)}_R(-\gamma,\beta|v)_{k',l'}
\label{proof2r}
\end{eqnarray}
and use the identity derived from 15.4.28 of \cite{baxbook}
\begin{equation}
\Theta_{m}(v_1)H_m(v_2)+H_m(v_1)\Theta_m(v_2)
=f^{R}_{+}(v_1+v_2)g^{R}_{+}(v_1-v_2)
\label{baxid1}
\end{equation}
with 
\begin{eqnarray}
&&f^{R}_{+}(z)=2H_m(z/2)\Theta_m(z/2)/(H_m(K)\Theta_m(K))\\
&&g^{R}_{+}(z)=H_m(K+z/2)\Theta_m(K+z/2)\label{grpdef}
\end{eqnarray}
where we note that
\begin{eqnarray}
&&f_{+}^R(-z)=-e^{i\pi m_2 z/(2L\eta)}f_{+}^R(z)\\
&&g^{R}_{+}(-z)=g_{+}^{R}(z)\label{grp1}\\
&&g^R_{+}(z+4L\eta)=(-1)^{m_1}(-1)^{m_1m_2}g^R_{+}(z)\label{grp3}
\end{eqnarray}
and for $m_1$ and $m_2$ even
\begin{equation}
g_{+}^R(z+2L\eta)=(-1)^{m_1/2}(-1)^{m_1m_2/4}g_{+}^R(z)\label{grp4}
\end{equation}
where (\ref{grp1}) follows from (\ref{grpdef}). 
The proof of (\ref{grp3}) follows from (\ref{Hminus}),(\ref{more1}),(\ref{more2})
and (\ref{Thsh}).
The property (\ref{grp4}) follows from (\ref{grp3}) with $m_1\rightarrow m_1/2$ 
and $m_2\rightarrow m_2/2$.

The diagonal elements of $W^{(1)R}(\alpha,\beta|v',v)$ are
\begin{equation}
W^{(1)R}(\alpha,\beta|v',v)_{11,11} = f_{+}^{R}(v+v'+2k)g_{+}^{R}(v-v')
\tau'_{\alpha,0}\tau_{\beta,0}
\label{rproof8}
\end{equation}
\begin{equation}
W^{(1)R}(\alpha,\beta|v',v)_{1,L;1,L} 
= f_{+}^R(v+v'+2K+2L\eta)g_{+}^R(v-v'+2L\eta)
\tau'_{\alpha,0}
\tau_{\beta,L}
\label{rproof9}
\end{equation}
\begin{equation}
W^{(1)R}(\alpha,\beta|v',v)_{L,1;L,1}=
f_{+}^{R}(v+v'+2K-2L\eta)g_{+}^R(v-v'+2L\eta)
\tau'_{\alpha,L}\tau_{\beta,0}
\label{rproof10}
\end{equation}
\begin{equation}
W^{(1)R}(\alpha,\beta|v',v)_{L,L;L,L}=
f_{+}^R(v+v'+2K)g_{+}^R(v-v'+4L\eta)
\tau'_{\alpha,L}\tau_{\beta,L}
\label{rproof11}
\end{equation}

The equalities
\begin{eqnarray}
&&W^{(1)R}(\alpha,\beta|v',v)_{1,1;1,1}
=W^{(1)R}(\alpha,\beta|v,v')_{1,1;1,1}\label{proofr15}\\
&&W^{(1)R}(\alpha,\beta|v',v)_{L,L;L,L}=W^{(1)R}(\alpha,\beta|v,v')_{L,L;L,L}
\label{proofr16}
\end{eqnarray}
follow from (\ref{grp3}).
and
\begin{eqnarray}
&&W^{(1)R}(\alpha,\beta|v',v)_{1,L;1,L}
=(-1)^{m_1m_2}(-1)^{m_1}W^{(1)R}(\alpha,\beta|v,v')_{1,L;1,L}
\label{proofr21}\\
&&W^{(1)R}(\alpha,\beta|v',v)_{L,1;L,1}
=(-1)^{m_1m_2}(-1)^{m_1}W^{(1)R}(\alpha,\beta|v,v')_{L,1;L,1}
\label{proofr22}
\end{eqnarray}
follow from (\ref{grp3}). Consequently with the restriction that $m_1$ 
is even or that both $m_1$ and $m_2$ are odd the methods of appendix C
of \cite{bax72} demonstrate that the interchange\ relation (\ref{commlra})
holds for $A=R$.

\subsubsection{The case $A=RS$}

For the case $A=RS$ we consider
\begin{equation}
Q_L^{(1)}(v')RSQ_R^{(1)}(v)|_{\alpha,\beta}
={\rm Tr}W^{(1)RS}(\alpha_1,\beta_1|v',v)
\cdots W^{(1)RS}(\alpha_N,\beta_N|v',v)
\label{proof1rs}
\end{equation} 
where $W^{(1)RS}(\alpha,\beta|v',v)$ are $L^2\times L^2$ matrices with elements
\begin{eqnarray}
&&W^{(1)RS}(\alpha,\beta|v',v)_{k,k';l,l'}=\sum_{\gamma=\pm}
\gamma S^{(1)}_L(\alpha,\gamma|v')_{k,l}
S^{(1)}_R(-\gamma,\beta|v)_{k',l'}
\label{proof2rs}
\end{eqnarray}
and use the identity
\begin{equation}
\Theta_m(v_1)H_m(v_2)-H_m(v_1)\Theta_m(v_2)
=f^{R}_{-}(v_1+v_2)g^{R}_{-}(v_1-v_2))
\label{baxid2}
\end{equation}
with 
\begin{eqnarray}
&&f^{R}_{-}(z)=2e^{-i\pi
    m_2K/2L\eta}H_m(z/2-K)\Theta_m(-z/2+K)/(H_m(K)\Theta_m(K)\\
&&g^{R}_{-}(z)=H_m(z/2)\Theta_m(-z/2)\label{grmdef}
\end{eqnarray}
where we note 
\begin{eqnarray}
&&g^{R}_{-}(-z)=-g^{R}_{-}(z)
\label{grm1}\\
&&g^R_{-}(z+4L\eta)=(-1)^{m_1+m_2}(-1)^{m_1m_2}g_{-}^R(z)\label{grm3}
\end{eqnarray}
and for both $m_1$ and $m_2$ even 
\begin{equation}
g_{-}^R(z+2L\eta)=(-1)^{(m_1+m_2)/2}(-1)^{m_1m_2/4}g_{-}^R(z)
\label{grm4}
\end{equation}
The relation (\ref{grm1}) follows from (\ref{Hminus}), the relation
(\ref{grm4}) follows from (\ref{grm3}) with $m_1\rightarrow
m_1/2,~m_2\rightarrow m_2/2$, and
the proof of (\ref{grm3}) follows from (\ref{Hminus}),(\ref{more1}),(\ref{more2}) and (\ref{Thsh}).
The diagonal elements of $W^{(1)R}(\alpha,\beta|v',v)$ are
\begin{equation}
W^{(1)RS}(\alpha,\beta|v',v)_{11,11} = f_{-}^R(v+v'+2K)g^R_{-}(v-v')
\tau'_{\alpha,0}\tau_{\beta,0}
\label{proof8rs}
\end{equation}
\begin{equation}
W^{(1)RS}(\alpha,\beta|v',v)_{1,L;1,L} 
= f_{+}^R(v+v'+2K+2L\eta)g_{-}^R(v-v'+2L\eta)\tau'_{\alpha,0}
\tau_{\beta,L}
\label{proof9rs}
\end{equation}
\begin{equation}
W^{(1)RS}(\alpha,\beta|v',v)_{L,1;L,1}=f_{-}(v+v'-2L\eta)g_{-}^R(v-v'+2L\eta)
\tau'_{\alpha,L}\tau_{\beta,0}
\label{proof10rs}
\end{equation}
\begin{equation}
W^{(1)RS}(\alpha,\beta|v',v)_{L,L;L,L}=f_{-}^R(v+v'+2K)g^R_{-}(v-v'+4L\eta)
\tau'_{\alpha,L}\tau_{\beta,L}
\label{proof11rs}
\end{equation}

When $N$ is odd the antisymmetry of $g_{-}^R(z)$ in (\ref{proof8rs})
prevents the proof of \cite{bax72} from being used. However when $N$ is even 
the operators $S$ and $R$ commute and thus (\ref{commlra}) with
  $A=RS$ holds if the equivalent relation
\begin{equation}
Q_L(v_1)RSQ_R(v_2)=Q_L(v_2)SRQ_R(v_1)
\label{rssr}
\end{equation}
is valid.
But 
\begin{equation}
Q_L^{(1)}(v')SRQ_R^{(1)}(v)|_{\alpha,\beta}
={\rm Tr}W^{(1)SR}(\alpha_1,\beta_1|v',v)
\cdots W^{(1)SR}(\alpha_N,\beta_N|v',v)
\label{proof1sr}
\end{equation} 
where 
\begin{eqnarray}
&&W^{(1)SR}(\alpha,\beta|v',v)_{k,k';l,l'}=\sum_{\gamma=\pm}
(-\gamma) S^{(1)}_L(\alpha,\gamma|v')_{k,l}
RS^{(1)}_R(-\gamma,\beta|v)_{k',l'}\nonumber\\
&&=-W^{(1)SR}(\alpha,\beta|v',v)_{k,k';l,l'}
\label{proof2sr}
\end{eqnarray}
and the minus sign in (\ref{proof2sr}) compensates for the
antisymmetry of $g_{-}^R(z)$. Therefore by use of
(\ref{grm3}) in the case
\begin{equation}
(-1)^{m_1+m_2}(-1)^{m_1m_2}=1
\label{rscondition}
\end{equation}
we find that (\ref{rssr}) holds and hence we find that the 
interchange relation
(\ref{commlra}) holds for $N$ even in the case $m_1$ and $m_2$ even
but fails in the other three cases.

\subsubsection{Summary}

The results obtained above for the validity of the interchange
relation (\ref{commlra}) for $A=I,S$ and $R$ are summarized in the
following table where Y (N) indicates that the relation holds (fails).

\begin{table}[h!]
\center
\caption{Summary of the values of the matrix $A$ for which the
interchange relation (\ref{commlra}) holds.}
\label{tab:1}
\begin{tabular}{|cc|cccc|}\hline
$m_1$&$m_2$&$I$&$S$&$R$&$RS$\\\hline
o&e&Y&Y&N&N\\
o&o&N&Y&Y&N\\
e&o&Y&N&Y&N\\
e&e&Y&Y&Y&Y\\\hline
\end{tabular}
\end{table}

\subsection{The nonsingularity  condition}

It remains to examine the validity of the nonsingularity condition. 
In the case of $m_2=0$ this condition was numerically 
studied in ref. \cite{fm1} for several values of $L$ and $N$ and it was found
that for $L$ odd that $Q_R(v)$ was nonsingular for all $v$ only when
$m_1$ was odd. We have extended that study to $m_2\neq 0$  and found
that for the cases studied $Q_R(v)$ is singular for all $v$ only when
$L$ is odd and both $m_1$ and $m_2$ are even.\\
In the remaining three cases where one or both of $m_1$ and $m_2$ are
odd the matrices $Q_R(v)$ and $Q_L(v)$ were non singular for generic
values of $v$, 
 We conjecture that this is generally true.

\subsection{The matrices $Q^{(1)}_{72}(v)$}

Using the results for the interchange  relation summarized in table 1
and assuming the validity of the conjecture of sec. 5.2 
on the nonsingularity of $Q_R^{(v)}$
we conclude that the
matrix $Q^{(1)}_{72}(v)$ defined by (\ref{q72}) satisfies the TQ equation
(\ref{tq1}) and the commutation relations (\ref{tq2}) and (\ref{tq3})
for both case 1 where $m_1$ is odd and $m_2$ is even and case 4 where
$m_1$ is even and $m_2$ is odd. It further follows from the 
relations summarized in table 1 that the commutation relations 
(\ref{c1})-(\ref{c4}) with R,
S, and RS hold for the three case where $m_1$ and $m_2$ are not both even..

\section{The matrix $Q^{(2)}_{72ee}(v;t)$ for case 4 where  $m_1,m_2$ is 
even and $N$ is even}

We found in sec.5.2 that in  case 4 where $L$ is odd 
and $m_1,m_2$ is even the matrix $Q^{(1)}_{72ee}$ does
not exist. Therefore to satisfy the TQ equation  
\begin{equation}
T(v)Q^{(2)}(v;t)=\omega^N[h(v+\eta)]^NQ^{(2)}(v-2\eta;t)
+\omega^{-N}[h(v-\eta)]^NQ^{(2)}(v+2\eta;t)
\label{n2tq1}
\end{equation}
and the
commutation relations (\ref{tq2})-(\ref{tq3}) a new construction must
be found. For $m_2=0$ this was accomplished in ref.\cite{klaus}. We
here generalize this construction to even values of $m_2\neq 0.$ 
In sec. 4.2 we demonstrated that the matrix $Q^{(2)}_R(v)$ defined by
the matrices $S^{(2)}_R(v;t)$ of (\ref{SRm2}) 
satisfies the equation (\ref{2tq1}) and that there is a
companion equation for $Q^{(2)}_L(v)$. 
Therefore to complete the proof of the TQ equation (\ref{n2tq1}) we
must find values of $t$ and matrices  $A$ for which 
\begin{equation}
Q^{(2)}_L(v';t)AQ^{(2)}_R(v;t)=Q_L^{(2)}(v;t)AQ_R^{(2)}(v';t)\label{q2comm}
\end{equation}
holds for which $Q_R^{(2)}(v;t)$ is nonsingular.

\subsection{The  interchange relations}

We consider the cases of $A=I,S,R$ and $RS$ separately.

\subsubsection{The case $A=I$}

We begin by examining (\ref{q2comm}) with $A=I$ and write
\begin{equation}
Q_L^{(2)}(v';t)Q_R^{(2)}(v;t)|_{\alpha,\beta}
={\rm Tr}W^{(2)}(\alpha_1,\beta_1|v',v)\cdots W^{(2)}(\alpha_N,\beta_N|v',v)
\label{2tq12}
\end{equation} 
where $W^{(2)}(\alpha,\beta|v',v)$ are $L^2\times L^2$ matrices with elements
\begin{equation}
W^{(2)}(\alpha,\beta|v',v)_{k,k';l,l'}=\sum_{\gamma=\pm}
S^{(2)}_L(\alpha,\gamma|v')_{k,l}
S^{(2)}_R(\gamma,\beta|v)_{k',l'}
\label{2tq13}
\end{equation}
The matrix $W^{(2)}(\alpha,\beta|v',v))$ is explicitly written out as
\begin{eqnarray}
&&W^{(2)}_{k,k';k+1,k'+1}(\alpha,\beta|v',v)
=\tau'_{\alpha,-k}\tau_{\beta,-k'}f_{-}(v'+v+2(k-k')\eta)
g_{-}(v'-v+2t+2(k+k')\eta)
\label{elem1}\\
&&W^{(2)}_{k+1,k'+1;k,k'}(\alpha,\beta|v',v)
=\tau'_{\alpha,k}\tau_{\beta,k'}f_{-}(v'+v-2(k-k')\eta)g_{-}
(v'-v-2t-2(k+k')\eta)
\label{elem2}\\
&&W^{(2)}_{k,k'+1;k+1,k'}(\alpha,\beta|v',v)
=\tau'_{\alpha,-k}\tau_{\beta,k'}f_{+}(v'+v+2t+2(k+k')\eta)g_{+}(v'-v+2(k-k')\eta)
\label{elem3}\\
&&W^{(2)}_{k+1,k';k,k'+1}(\alpha,\beta|v',v)=\tau'_{\alpha,k}\tau_{\beta,-k'}
f_{+}(v'+v-2t-2(k+k')\eta)g_{+}(v'-v-2(k-k')\eta)
\label{elem4}
\end{eqnarray}
where $f_{+}(z)$ and $g_{+}(z)$ are given by (\ref{proof4}) and (\ref{proof5})
and $f_{-}(z)$ and $g_{-}(z)$ are given by (\ref{proofs6}) and (\ref{proofs7}).

We again look for an $L^2\times L^2$ diagonal matrix $Y^{(2)}$ 
\begin{equation}
Y_{m,m';k,k'}=y_{m,m'}\delta_{m,k}\delta_{m',k'}
\label{ydiag}
\end{equation}
such that
\begin{equation}
W^{(2)}(\alpha,\beta|v',v)=Y^{(2)}W^{(2)}(\alpha,\beta|v,v')Y^{(2)-1}
\label{2tq14}
\end{equation}
The expressions for the diagonal elements are  symmetric under the interchange
of $v'$ and $v$ and thus there are no restrictions such as we had for
$Q^{(1)}_{72}$ and $Q^{(1)S}_{72}$. Using (\ref{elem1})-(\ref{elem2})
and  $g_{-}(z)=g_{-}(-z)$ in (\ref{ydiag})-(\ref{2tq14})  we find the
single equation
\begin{equation}
g_{-}(v'-v+2t+2(k+k')\eta)=\frac{y_{k,k'}}{y_{k+1,k'+1}}
g_{-}(v-v'+2t+2(k+k')\eta)\label{elem5}
\end{equation}
or equivalently
\begin{equation}
y_{k+1,k'+1}=y_{k,k'}\frac{g_{-}(v'-v-2(k+k')\eta-2t)}
{g_{-}(v'-v+2(k+k')\eta+2t)}
\label{elem6}
\end{equation}
Similarly by using (\ref{elem3})-(\ref{elem4}) and $g_{+}(z)=g_{+}(-z)$ in 
(\ref{ydiag})-(\ref{2tq14})  we find the
single equation
\begin{equation}
g_{+}(v'-v+2(k-k')\eta)=\frac{y_{k,k'+1}}{y_{k+1,k'}}g_{+}(v-v'+2(k-k')\eta)
\label{elem7}
\end{equation}
and thus we obtain
\begin{equation}
y_{k+1,k'-1}=y_{k,k'}\frac{g_{+}(v-v'+2(k-k'+1)\eta)}
{g_{+}(v-v'-2(k-k'+1)\eta)}
\label{elem8}
\end{equation}
We follow \cite{klaus} and note that for  the recursions (\ref{elem6})
to be 
free of contradictions we need
\begin{equation}
y_{k+L,k'+L}=y_{k,k'}
\label{elem9} 
\end{equation}
and for (\ref{elem8}) to be free of contradictions 
\begin{equation}
y_{k-L,k'+L}=y_{k,k'}
\label{elem10}
\end{equation}

In order for (\ref{elem9}) to hold we need to choose $t$ such that
\begin{eqnarray}
&&\frac{g_{-}(v-v'-2(k+k')\eta-4(L-1)\eta-2t)}
{g_{-}(v-v'+2(k+k')\eta+4(L-1)\eta+2t)}
\frac{g_{-}(v-v'-2(k+k')\eta-4(L-2)\eta-2t)}
{g_{-}(v-v'+2(k+k')\eta+4(L-2)\eta+2t)}\nonumber\\
&&\frac{g_{-}(v-v'-2(k+k')\eta-2t)}
{g_{-}(v-v'+2(k+k')\eta+2t)}=1
\label{final10}
\end{eqnarray}
which will be satisfied if the factors 
$g_{-}(v-v'-2(k+k')\eta-4c_1\eta-2t)$ in the numerator must   
cancel the factors  $g_{-}(v-v'+2(k+k')\eta+4c_2\eta+2t)$
in the denominator. For this we need to use the periodicity properties of
$g_{-}(v).$

From the definition (\ref{proofs7}) of $g_{-}(v)$
 and the periodicity of $H_m(v)$ (\ref{hm1}),(\ref{perioddef})
it follows that for all even $m_1$ and $m_2$
\begin{equation}
g_{-}(v+4(r_1K+ir_2K'))=g_{-}(v)
\label{gsperiod10}
\end{equation}
Furthermore we use the definitions (\ref{tq4}) and (\ref{rdefs})  in 
(\ref{newgmprop}) to
find
\begin{equation}
g_{-}(z+r_0(r_1K+ir_2K'))=(-1)^{m_1m_2/4}(-1)^{m_2/2}g_{-}(z).
\label{gsperiod11}
\end{equation}
When $m_2\equiv 2 ({\rm mod}4)$ we see from (\ref{rdefs}) that
$r_0\equiv 2~({\rm mod}4)$ and thus it follows from
(\ref{gsperiod10}) and (\ref{gsperiod11}) that we have the additional
periodicity condition
\begin{equation}
g_{-}(z+2(r_1K+ir_2K'))=-(-1)^{m_1/2}g_{-}(z)
\label{gsperiod12}
\end{equation}

Consider first  the periodicity (\ref{gsperiod10}).
This will provide cancellation if an integer $I$ can be found such that
\begin{equation}
-2(k+k')\eta-4c_1\eta-2t=2(k+k')\eta+4c_2\eta+2t+4I(r_1K+ir_2K')
\label{gsproof11}
\end{equation}
which by multiplying by $L$, using (\ref{tq4}) and (\ref{rdefs})  
and defining 
\begin{equation}
t={\bar t}\eta
\label{tb}
\end{equation}
becomes
\begin{equation}
-(k+k')r_0-c_1r_0-{\bar t}r_0=c_2r_0+2LI
\label{gsproof12}
\end{equation}
For $m_2$ even the quantity $r_0$ is always even and thus
(\ref{gsproof12}) can always be satisfied by integers for ${\bar
  t}=n$ with $n$ an integer because $c_2$ can be shifted 
into the interval $0 \leq c_2 < L$. 
Furthermore if $m_2\equiv 0~({\rm mod}4)$ then for even $m_1$
we have $r_0\equiv 0~({\rm mod}4)$ and thus (\ref{gsproof12}) may be
satisfied by integers for ${\bar t}=n+1/2$. Thus we have demonstrated
that for all cases of $m_1$ and $m_2$ even that (\ref{elem9}) holds
for ${\bar t}=n$ and for $m_2\equiv 0~({\rm mod}4)$ and $m_1$ even that
(\ref{q2comm}) with $A=1$ is satisfied for ${\bar t}=n+1/2$ 
but is not satisfied if $m_2\equiv 2~({\rm mod}4)$ and $m_1$ even that
(\ref{q2comm}) with $A=1$ is satisfied for ${\bar t}=n+1/2$ but is not
satisfied if $m_2\equiv 2 ({\rm mod}4)$.

We next consider the periodicity condition (\ref{gsperiod12}) which
holds for $m_2\equiv 2~({\rm mod 4})$ and $r_0\equiv 2~({\rm mod}4)$
which with the additional restriction that $m_1\equiv 2~({\rm mod}4)$
specializes to
\begin{equation}
g_{-}(z+2(r_1K+ir_2K'))=g_{-}(z)
\label{gsproof13}
\end{equation}
This will give the desired cancellation in (\ref{final10}) if instead of
(\ref{gsproof12}) we have
\begin{equation}
-(k+k')r_0-c_1r_0-{\bar t}r_0=c_2r_0+LI
\label{gsproof14}
\end{equation}
Using the fact that  $r_0\equiv 2~({\rm mod}4)$
we see that this equation can be satisfied in integers for ${\bar t}=n+1/2$.
Thus we have demonstrated that (\ref{elem9}) is satisfied for the
case $m_1\equiv 2~({\rm mod}4$ and $m_2\equiv 2~({\rm mod}4)$ with 
$t=(n+1/2)\eta$.

To complete the proof of (\ref{q2comm}) with $A=1$ it remains to demonstrate
that (\ref{elem10}) holds. We see from (\ref{elem8}) that this will be
the case  
\begin{eqnarray}
&&\frac{g_{+}(v-v'+2(k-k'+1)\eta-4(L-1)\eta)}
{g_{+}(v-v'-2(k-k'+1)\eta+4(L-1)\eta)}
\frac{g_{+}(v-v'+2(k-k'+1)\eta-4(L-2)\eta)}
{g_{+}(v-v'-2(k-k'+1)\eta+4(L-2)\eta)}\nonumber\\
&&\frac{g_{+}(v-v'+2(k-k'+1)\eta)}
{g_{+}(v-v'-2(k-k'+1)\eta)}=1
\label{gsproof15}
\end{eqnarray}
In contrast to (\ref{final10}) this is independent of $t$ and
therefore using the periodicity condition which follows from the
definition (\ref{proof5}) of $g_{+}(z)$ and the periodicity of  
$H_m(v)$ (\ref{hm1}),(\ref{perioddef})
\begin{equation}
g_{+}(v+4(r_1K+ir_2K'))=g_{+}(v)
\label{gsproof16}
\end{equation}
we show that (\ref{gsproof15}) holds for all $m_1$ and $m_2$ even.
Thus we have proven that (\ref{q2comm}) with $A=I$ holds for the cases
\begin{eqnarray}
&&m_1~{\rm even},~~m_2~{\rm even~~with}~~t=n\eta\\
&&m_1~{\rm even},~~m_2\equiv 0~({\rm mod}4)~~{\rm with}~~t=(n+1/2)\eta\\
&&m_1\equiv 2~({\rm mod}4),~~m_2\equiv 2~({\rm mod}4)~~{\rm
    with}~~t=(n+1/2)\eta
\end{eqnarray}

\subsubsection{The case $A=S$}

We study the interchange relation (\ref{q2comm}) with $A=S$ in a completely
analogous manner by writing
\begin{equation}
Q_L^{(2)}(v';t)SQ_R^{(2)}(v;t)|_{\alpha,\beta}
={\rm Tr}W^{(2S)}(\alpha_1,\beta_1|v',v)\cdots W^{(2S)}(\alpha_N,\beta_N|v',v)
\label{gsproof17}
\end{equation} 
where $W^{(2S)}(\alpha,\beta|v',v)$ are $L^2\times L^2$ matrices with elements
\begin{equation}
W^{(2S)}(\alpha,\beta|v',v)_{k,k';l,l'}=\sum_{\gamma=\pm}
S^{(2)}_L(\alpha,\gamma|v')_{k,l}\gamma
S^{(2)}_R(\gamma,\beta|v)_{k',l'}
\label{gspeoof18}
\end{equation}
is explicitly written out as
\begin{eqnarray}
&&W^{(2S)}_{k,k';k+1,k'+1}(\alpha,\beta|v',v)
=\tau'_{\alpha,-k}\tau_{\beta,-k'}f_{+}(v'+v+2(k-k')\eta)g_{+}
(v'-v+2t+2(k+k')\eta)\\
&&W^{(2S)}_{k+1,k'+1;k,k'}(\alpha,\beta|v',v)
=\tau'_{\alpha,k}\tau_{\beta,k'}f_{+}(v'+v-2(k-k')\eta)g_{+}
(v'-v-2t-2(k+k')\eta)\\
&&W^{(2S)}_{k,k'+1;k+1,k'}(\alpha,\beta|v',v)
=\tau'_{\alpha,-k}\tau_{\beta,k'}f_{-}(v'+v+2t+2(k+k')\eta)
g_{-}(v'-v+2(k-k')\eta)\\
&&W^{(2S)}_{k+1,k';k,k'+1}(\alpha,\beta|v',v)=\tau'_{\alpha,k}\tau_{\beta,-k'}
f_{-}(v'+v-2t-2(k+k')\eta)g_{-}(v'-v-2(k-k')\eta)
\end{eqnarray}
where the roles of  $f_{+}(z)$ and $g_{+}(z)$ 
and $f_{-}(z)$ and $g_{-}(z)$ are reversed from (\ref{elem1})-(\ref{elem4}).

We now use the periodicity property
\begin{equation}
g_{+}(z+r_0(r_1K+ir_2K'))=(-1)^{m_1m_2/4}g_{+}(z)
\end{equation}
which follows from(\ref{newgpprop}) and (\ref{gsproof16}) in exactly
the same manner used to prove (\ref{q2comm}) with $A=I$ to prove that
(\ref{q2comm}) with $A=S$
\begin{eqnarray}
&&m_1~{\rm even},~~m_2~{\rm even~~with}~~t=n\eta\\
&&m_1~{\rm even},~~m_2\equiv 0~({\rm mod}4)~~{\rm with}~~t=(n+1/2)\eta\\
&&m_1\equiv 0~({\rm mod}4),~~m_2\equiv 2~({\rm mod}4)~~{\rm
    with}~~t=(n+1/2)\eta
\end{eqnarray}

\subsubsection{The case $A=R$}

To investigate (\ref{q2comm}) with $A=R$ 
we write
\begin{equation}
Q_L^{(2)}(v';t)RQ_R^{(2)}(v;t)|_{\alpha,\beta}
={\rm Tr}W^{(2R)}(\alpha_1,\beta_1|v',v)\cdots W^{(2R)}(\alpha_N,\beta_N|v',v)
\label{grproof17}
\end{equation} 
where $W^{(2R)}(\alpha,\beta|v',v)$ are $L^2\times L^2$ matrices with elements
\begin{equation}
W^{(2R)}(\alpha,\beta|v',v)_{k,k';l,l'}=
\sum_{\gamma=\pm}
S^{(2)}_L(\alpha,\gamma|v')_{k,l}
S^{(2)}_R(-\gamma,\beta|v)_{k',l'}
\label{grproof1}
\end{equation}

By use of the identities (\ref{baxid1}) and (\ref{baxid2}) the matrix
$W^{(2)R}(\alpha,\beta|v',v)$ is explicitly written out as 
\begin{eqnarray}
&&W^{(2R)}_{k,k';k+1,k'+1}(\alpha,\beta|v',v)\nonumber\\
&&=\tau'_{\alpha,-k}\tau_{\beta,-k'}
f_{-}^{R}(v+v'+2(k-k')\eta)g_{-}^{R}(v'-v+2t+2(k+k')\eta)\label{1grproof2}\\
&&W^{(2R)}_{k+1,k'+1;k,k'}(\alpha,\beta|v',v)\nonumber\\
&&=\tau'_{\alpha,k}\tau_{\beta,k'}
f_{-}^{R}(v+v'-2(k-k')\eta)g_{-}^{R}(v'-v-2t-2(k+k')\eta)\label{2grproof2}\\
&&W^{(2R)}_{k,k'+1;k+1,k'}(\alpha,\beta|v',v)\nonumber\\
&&=\tau'_{\alpha,-k}\tau_{\beta,k'}
f_{+}^{R}(v+v'+2t+2(k+k')\eta)g_{+}^{R}(v'-v-2(k-k')\eta)\label{3grproof2}\\
&&W^{(2R)}_{k+1,k';k,k'+1}(\alpha,\beta|v',v)\nonumber\\
&&=-\tau'_{\alpha,k}\tau_{\beta,-k'}
f_{+}^{R}(v+v'-2t-2(k+k')\eta)g_{+}^{R}(v'-v-2(k-k')\eta)
\label{4grproof2}
\end{eqnarray}

We again follow the procedure of the previous subsection and look for a
diagonal similarity transformation.
However because of the antisymmetry of $g_{-}^{R}(z)$ 
and recalling the $L$ is odd 
the consistency condition analogous to (\ref{final10}) 
for (\ref{1grproof2}) and (\ref{2grproof2}) is
\begin{eqnarray}
&&\frac{g^R_{-}(v-v'-2(k+k')\eta-4(L-1)\eta-2t)}
{g^R_{-}(v-v'+2(k+k')\eta+4(L-1)\eta+2t)}
\frac{g^R_{-}(v-v'-2(k+k')\eta-4(L-2)\eta-2t)}
{g^R_{-}(v-v'+2(k+k')\eta+4(L-2)\eta+2t)}\cdots\nonumber\\
&&\frac{g^R_{-}(v-v'-2(k+k')\eta-2t)}
{g^R_{-}(v-v'+2(k+k')\eta+2t)}=-1
\label{proofq2r1}
\end{eqnarray}
whereas in analogy with (\ref{gsproof15}) the consistency condition
for (\ref{3grproof2}) and (\ref{4grproof2}) is
\begin{eqnarray}
&&\frac{g^R_{+}(v-v'+2(k-k'+1)\eta-4(L-1)\eta)}
{g^R_{+}(v-v'-2(k-k'+1)\eta+4(L-1)\eta)}
\frac{g^R_{+}(v-v'+2(k-k'+1)\eta-4(L-2)\eta)}
{g^R_{+}(v-v'-2(k-k'+1)\eta+4(L-2)\eta)}\cdots\nonumber\\
&&\frac{g_{+}^R(v-v'+2(k-k'+1)\eta)}
{g^R_{+}(v-v'-2(k-k'+1)\eta)}=1
\label{proofq2r2}
\end{eqnarray}
we see that we will satisfy (\ref{proofq2r1}) if
there is the pairwise relation between factors in the numerator and
denominator of
\begin{equation}
g_{-}^R(v-v'-2(k+k')\eta-4c_1\eta-2t)=-g_{-}^R(v-v'+2(k+k')\eta+4c_2\eta+2t)
\label{consist1}
\end{equation}
and therefore in contrast with the cases $A=I$ and $S$ we must
consider antiperiodicity properties of $g_{-}^{R}(z)$

In contrast the condition (\ref{proofq2r2}) is satisfied if
\begin{equation}
g_{+}^R(v-v'+2(k-k'+1)\eta-4c_1\eta)=g_{+}^R(v-v'-2(k-k'+1)\eta+4c_1\eta)
\label{consist2}
\end{equation}
which requires a periodicity (and not an antiperiodicity) property for
$g_{+}^R(z)$

The (anti)periodicity properties analogous to (\ref{gsperiod10})
follow from the definitions (\ref{grpdef}) of $g_{+}^R(z)$ and
(\ref{grmdef}) of $g^R_{-}(z)$ and the properties 
(\ref{hm1}) and (\ref{tm1})
of $H_m(z)$ and $\Theta_m(z)$
\begin{equation}
g_{\pm}^R(z+4(r_1K+ir_2K')=(-1)^{r_1}g_{\pm}^R(z)
\label{rperiods}
\end{equation}

Consider first the the condition (\ref{consist2}). In the case that 
$r_1$ is even we have from (\ref{rperiods}
\begin{equation}
g_{+}^R(z+4(r_1K+ir_2K'))=g^R_{+}(z)
\end{equation}
and thus(\ref{consist2}) will hold if
\begin{equation}
4(k-k'+1)\eta-4c_1=4c_2+4I(r_1K+ir_2K')
\end{equation}
which after we multiply by $L$ and use the root of unity condition (\ref{tq4})
becomes
\begin{equation}
(k-k'+1)r_0-c_1r_0=c_2r_0+2LI
\end{equation}
which can always be satisfied in integers because for $m_1$ and $m_2$
even $r_0$ is always even.

In the opposite case where $r_1$ is odd we have from (\ref{rperiods})
\begin{equation}
g_{+}^R(z+8(r_1K+ir_2K'))=g^R_{+}(z)
\end{equation}
and thus (\ref{consist2}) will be satisfied if
\begin{equation}
(k-k'+1)r_0-c_1r_0=c_2r_0+4LI
\end{equation}
This can only hold if $r_0\equiv 0 ({\rm mod}4)$. However it follows
from the definition (\ref{rdefs}) of $r_1$ that when  $m_1$ is even  and
$r_1$ is odd that $r_0/4$ must be an integer. Therefore the condition 
(\ref{consist2}) is always satisfied.

We next attempt to satisfy (\ref{consist1}) by use of 
(\ref{rperiods}) for $r_1$ odd. Thus
\begin{equation}
g_{-}^R(z+4(2I+1)(r_1K+ir_2K'))=-g_{-}^R(z)
\end{equation}
and (\ref{consist1}) will be satisfied is
\begin{equation}
-(k+k')r_0-c_1r_0-r_0{\bar t}=c_2r_0+2(2I+1)L
\label{consist3}
\end{equation}
However, we have just seen that when $r_1$ is odd and $m_1$ is even
that $r_0/4$ and $m_2/4$ must be integers. Therefore because $2I+1$ and $L$ are
odd (\ref{consist3}) cannot be satisfied in integers  for any integer ${\bar
  t}$. Therefore there are no solutions for $A=R$ which are analogous
to the solutions $t=n\eta$ for the cases $A=I$ and $S$.
However, if ${\bar t}=n+1/2$ the condition (\ref{consist3}) can be
satisfied in integers if $r_0/4$ is an odd integer which leads to the
conclusion that the  interchange relation (\ref{q2comm}) with $A=R$ holds
for $t=(n+1/2)\eta$ when
\begin{equation}
m_1\equiv 2~({\rm mod}4),~~~m_2\equiv 0~({\rm mod}4)
\end{equation} 

It remains to investigate the possibility of using (\ref{grm4}) to
satisfy (\ref{consist1}). The condition (\ref{grm4}) is antiperiodic
in the three cases where $m_1/4$ and $m_2/4$ are not both integers. We
have already considered the case $m_2\equiv 0~({\rm mod 4})$ and thus
need only consider the remaining cases where $m_2\equiv 2~({\rm mod
  4}).$ In these cases $r_0\equiv 2~({\rm mod}4)$ and thus, because
$m_1$ is even $r_1$ must be even. Thus we find from (\ref{rperiods})  
and (\ref{grm4}) that for $m_1$ even and $m_2\equiv 2~({\rm mod 4})$
that
\begin{equation}
g_{-}^{R}(z+2(2I+1)(r_1K+ir_2K'))=-g_{-}^R(z)
\end{equation}
and from this it follows that (\ref{proofq2r1}) holds for $t=(n+1/2)\eta$ 

\subsubsection{The case $A=RS$}

The case $A=RS$ is treated by similar methods and we find that the
only case where the interchange relation (\ref{q2comm}) holds is for
$t=(n+1/2)\eta$, $m_1\equiv 2~({\rm mod} 4)$ and $m_2\equiv 0~({\rm mod}4).$

\clearpage

\subsubsection{Summary}

The results obtained above for the validity of the interchange
relation (\ref{q2comm})  with $t=n\eta$ and $(n+1/2)\eta$ for $A=I,S$ and $R$ 
are summarized in the
following table where Y (N) indicates that the relation holds (fails).

\begin{table}[h!]
\center
\caption{Summary of the values of the matrix $A$ for which the
   interchange relation (\ref{q2comm}) with $t=n\eta$ holds where the
  notation $0 (2)$ stands for $\equiv 0(2) ({\rm mod}4)$}
\label{tab:2}
\begin{tabular}{|cc|cccc|}\hline
$m_1$&$m_2$&$I$&$S$&$R$&$RS$\\\hline
0&0&Y&Y&N&N\\
2&0&Y&Y&N&N\\
0&2&Y&Y&N&N\\
2&2&Y&Y&N&N\\\hline
\end{tabular}
\end{table}

\begin{table}[h!]
\center
\caption{Summary of the values of the matrix $A$ for which the
 interchange relation (\ref{q2comm}) with $t=(n+1/2)\eta$ holds where
  the notation
$0(2)$ stands for $\equiv 0(2)~({\rm mod}4)$}
\label{tab:3}
\begin{tabular}{|cc|cccc|}\hline
$m_1$&$m_2$&$I$&$S$&$R$&$RS$\\\hline
0&0&Y&Y&N&N\\
2&0&Y&Y&Y&Y\\
0&2&N&Y&Y&N\\
2&2&Y&N&Y&N\\\hline
\end{tabular}
\end{table}

\subsection{The nonsingularity condition}

We have found numerically for several special cases that 
$Q^{(2)}_R(v;n\eta)$ is non singular for $m_1$ and $m_2$ even  
and that $Q^{(2)}_R(v;(n+1/2)\eta)$ is singular only for $m_1\equiv
2~({\rm mod}4)$ and $m_2\equiv 0~({\rm mod}4).$ 
We conjecture that this is true generally.

\subsection{The matrices $Q_{72ee}^{(2)}(v;t)$ for $t=n\eta$ and $(n+1/2)\eta$}

From the results for the  interchange relation (\ref{q2comm}) summarized
in table 2 for $t=n\eta$ and in table 3 for $t=(n+1/2)\eta$ and
assuming the validity of the nonsingularity conjecture we conclude
that the matrix $Q^{(2)}_{72}(v;n\eta)$ is constructed from
(\ref{q72}) for all even $m_1$ and $m_2$ with the commutation relation
given in (\ref{c4a}) of the introduction and that the matrix
$Q^{(2)}_{72}(v;(n+1/2)\eta)$ is similarly constructed from
(\ref{q72}) for the three cases 
\begin{eqnarray}
&&m_1\equiv 0~({\rm mod}4),~~~m_2\equiv 0~({\rm mod}4)\nonumber\\
&&m_1\equiv 2~({\rm mod}4),~~~m_2\equiv 2~({\rm mod}4)\nonumber\\
&&m_1\equiv 0~({\rm mod}4),~~~m_2\equiv 2~({\rm mod}4)
\end{eqnarray}
with the commutation relations given by (\ref{c4b})-(\ref{c4d}).\\

\section{Quasiperiodicity properties of $Q^{(1)}_{72}(v)$ 
and $Q^{(2)}_{72}(v;t)$}

We complete our discussion of the $Q$ matrices by 
deriving their quasiperiodicity properties and general form of the
eigenvalues.

\subsection{Quasiperiodicity of $Q^{(1)}_{72}(v)$}

To compute the quasiperiodicity properties of $Q^{(1)}_{72}(v)$ we
first use the quasiperiodicity properties (\ref{hm1})-(\ref{tm2}) 
in the definition of $S_{R}^{(1)}$ (\ref{SRm}) to find
\begin{equation}
 S^{(1)}_R(\alpha,\beta)_{j,k}(v+\omega_1) =   
 (-\alpha)^{r_1}(-1)^{r_1r_2}S^{(1)}_R(\alpha,\beta)_{j,k}(v)
\label{srqp1}
\end{equation}
and
\begin{equation}
\begin{array}{lclcl} 
 S^{(1)}_R(\alpha,\beta)_{k,k+1}(v+\omega_2) 
& = &(-\alpha)^b(-1)^{ab}q'^{-1}e^{-2\pi iv/\omega_1}
e^{4\pi i k\eta/\omega_1}S^{(1)}_R(\alpha,\beta)_{k,k+1}(v)   
&~~&1\leq
 k\leq L-1\\
S^{(1)}_R(\alpha,\beta)_{k+1,k}(v+\omega_2) 
& = &  (-\alpha)^b(-1)^{ab}q'^{-1}e^{-2\pi iv/\omega_1}
e^{-4\pi i k\eta/\omega_1}S^{(1)}_R(\alpha,\beta)_{k+1,k}(v)   
&~~&1\leq k\leq L-1\\
S^{(1)}_R(\alpha,\beta)_{1,1}(v+\omega_2)   
& = & (-\alpha)^b(-1)^{ab}
q'^{-1}e^{-2\pi iv/\omega_1}S^{(1)}_R(\alpha,\beta)_{1,1}(v)&~~&\\
S^{(1)}_R(\alpha,\beta)_{L,L}(v+\omega_2)   
& = & (-\alpha)^b(-1)^{ab}(-1)^{r_0}q'^{-1}e^{-2\pi iv/\omega_1}
S^{(1)}_R(\alpha,\beta)_{L,L}(v)&~~&
\end{array}
\label{srqp2}
\end{equation}

The dependence of (\ref{srqp2}) on $r_0$ distinguishes case 1 with
$m_1$ odd and $m_2$ even from cases 2 and 3 with $m_2$ odd.

\subsubsection{Case 1 with $m_1$ odd and $m_2$ even}

In case 1 where $m_1$ is odd and $m_2$ is even the greatest common
factor $r_0$ in $2m_1$ and $m_2$ is even, $r_1$ must be
odd but $r_2$ is unrestricted. Therefore we find directly from the quasiperiodicity (\ref{srqp1})
and from (\ref{tqr0}) and (\ref{q72}) 
that 
\begin{equation}
Q^{(1)}_{72oe}(v+\omega_1)=-S(-1)^{Nr_2}Q^{(1)}_{72oe}(v)
\label{qpoe1}
\end{equation}

To examine quasiperiodicity under $v\rightarrow v+\omega_2$ we use the
diagonal similarity transformation (2.13) of \cite{fm1} to write
(\ref{srqp2}) as
\begin{equation}
S^{(1)}_R(\alpha,\beta)(v+\omega_2)
=(-\alpha)^b(-1)^{ab}q'^{-1}e^{-2\pi i v/\omega_1}
M^{(1)}S^{(1)}_R(\alpha,\beta)(v)M^{(1)-1}
\label{srqp2m}
\end{equation}
with
\begin{equation}
M^{(1)}_{k,k'}=e^{-2\pi i \eta k(k-1)/\omega_1}\delta_{k,k'}
\label{similarity}
\end{equation}
Thus we find from (\ref{tqr0}) and (\ref{q72}) that
\begin{equation}
Q^{(1)}_{72oe}(v+\omega_2)=(-S)^{b}(-1)^{Nab}e^{-2\pi iNv/\omega_1}
q'^{-N}Q^{(1)}_{72oe}(v)
\label{qpoe2}
\end{equation}
It follows from (\ref{qpoe1}), (\ref{qpoe2}) and the fact that the
eigenvectors of $Q^{(1)}_{72}(v)$ are independent of $v$ that
$Q^{(1)}_{72oe}(v)$ commutes with $S$ (as was shown directly in sec. 5)

It follows from (\ref{qpoe1}) and (\ref{qpoe2}) that $Q^{(1)}_{72oe}(v)$
has $N$ zeros in the
fundamental parallelogram (\ref{funp})
\begin{equation}
0,~\omega_1,~\omega_1+\omega_2,~\omega_2
\label{paroe}
\end{equation}
and thus may be written in factorized form as
\begin{equation}
Q^{(1)}_{72oe}(v)
={\cal K}{\rm exp}(-i\nu \pi
v/\omega_1)\prod_{j=1}^{N}H_m(v-v_j)
\label{form2}
\end{equation}
with $v_j$ in the parallelogram (\ref{paroe}).
Using the form (\ref{form2}) we find from the quasiperiodicity condition
(\ref{qpoe1}) 
\begin{equation}
1=e^{\pi i(1+\nu_S+\nu+N)}
\end{equation}
and thus
\begin{equation}
1+\nu+N+\nu_S={\rm even~ integer}
\label{sum1}
\end{equation}
where $(-1)^{\nu_S}$ is the eigenvalue of $S$.

From (\ref{qpoe2}) we find
\begin{equation}
(-S)^b=e^{-\nu \pi i \omega_2/\omega_1}(-1)^{bN}
{\rm exp}(2\pi i\sum_{j=1}^N(v_j+K)/\omega_1)
\end{equation}
and thus
\begin{equation}
b(\nu_S+1) -\nu \omega_2/\omega_1+bN+2\sum_{j=1}^N(v_j+K)/\omega_1
={\rm even~integer}
\label{sum2}
\end{equation}

\subsubsection{Cases 2 and 3 with $m_2$ odd}

When  $m_2$  is odd 
we see from (\ref{rdefs}) that $r_0$ and $r_2$ are odd and  $r_1$
is even for both $m_1$ even and odd.
Therefore because $r_0$ is odd we find instead
of the quasiperiodicity condition (\ref{srqp2m}) under $v\rightarrow
v+\omega_2$ we have instead a quasiperiodicity under $v\rightarrow v+2\omega_2$
\begin{equation}
S^{(1)}_R(\alpha,\beta)(v+2\omega_2)
=q'^{-4}e^{-4\pi i v/\omega_1}M^{(1)2}S^{(1)}_R(\alpha,\beta)(v)M^{(1)-2}
\label{more100}
\end{equation}
to find
\begin{equation}
Q^{(1)}_R(v+2\omega_2)=q'^{-4N}e^{-4\pi i Nv/\omega_1}Q^{(1)}_R(v)
\label{more101}
\end{equation}
Thus by use of (\ref{q72}) we find
\begin{equation}
Q^{(1)}_{72xo}(v+2\omega_2) = q'^{-4N}e^{-4\pi i Nv/\omega_1}Q^{(1)}_{72xo}(v)
\label{qpeo2}
\end{equation}
where $x$ is either e or o.

If the area of the fundamental parallelogram is to be $4KK'$ then the
quasiperiodic property (\ref{qpeo2}) mandates that instead of the
parallelogram (\ref{paroe}) we need to consider the parallelogram
\begin{equation}
0,~\omega_1/2,\omega_1/2+2\omega_2,~2\omega_2
\label{paraeo}
\end{equation}
To obtain the periodicity properties under $v\rightarrow \omega_1/2$
we write
\begin{equation}
\omega_1/2=r_1K+ir_2K'
\end{equation}
where $r_1/2$ is an integer because $r_1$ is even.
It then follows from the definitions (\ref{Hm}) and the properties
(\ref{app6}),(\ref{more1}) and (\ref{more2}) 
that
\begin{eqnarray}
&&H_m(v+\omega_1/2)=(-1)^{r_1/2}e^{\pi i r_1r_2/4}\Theta_m(v)\label{halfh}\\
&&\Theta_m(v+\omega_1/2)=e^{\pi i r_1r_2/4}H_m(v)
\label{half}
\end{eqnarray}
We therefore obtain for  $m_2$ odd and all $m_1$ that
\begin{equation}
S^{(1)}_R(v+\omega_1/2)
=e^{\pi i r_1r_2/4}RS^{r_1/2}S_R^{(1)}(v)
\label{om2qp}
\end{equation}

{\bf Case 3 with $m_1$ even}

\vspace{.1in}

When $m_1$ is further restricted to be even we find from (\ref{rdefs})
that $r_1/2$ is even and therefore (\ref{om2qp}) may be written
as
\begin{equation}
S_R^{(1)}(v+\omega_1/2)=e^{\pi i r_1r_2/4}RS_R^{(1)}(v)
\end{equation}
Therefore we find from (\ref{tqr0}) and (\ref{q72}) that
\begin{equation}
Q_{72eo}^{(1)}(v+\omega_1/2)=e^{N\pi i r_1r_2/4}RQ_{72eo}^{(1)}(v)
\label{qpeo1}
\end{equation}
and from (\ref{qpeo1}) and the fact that the eigenvectors of
$Q_{72eo}^{(1)}(v)$ are independent of $v$  it follows that
\begin{equation}
[Q^{(1)}_{72eo}(v),R]=0
\end{equation}
which has been directly proven in sec. 5.

From the quasiperiodicity relations (\ref{qpeo2}) and (\ref{qpeo1}) it
follows that the eigenvalues of $Q^{(1)}_{72eo}(v)$ may be written in
terms of $H_m(v)$ as
\begin{equation}
Q^{(1)}_{72eo}(v)={\cal K}e^{-\nu 2\pi i v/\omega_1}
\prod_{j=1}^NH_m(v/2-v_j/2)H_m(v/2-v_j/2+\omega_1/4)
H_m(v/2-v_j/2+\omega_1/2)H_m(v/2-v_j/2+3\omega_1/4)
\label{formeo}
\end{equation}
where the $N$ roots $v_j$ lie in the fundamental parallelogram (\ref{paraeo}).
From the quasiperiodicity relation (\ref{qpeo1}) we find the sum rule
\begin{equation}
e^{N\pi i r_1r_2/4}(-1)^{\nu_R}=(-1)^\nu
\label{sumeo1}
\end{equation}
where $(-1)^{\nu_R}$ are the eigenvalues of $R$ 
and from the quasiperiodicity relation (\ref{qpeo2}) we find the sum
rule
\begin{equation}
1={\rm exp}\left(-\nu 4\pi i\omega_2/\omega_1
+4\pi i\sum_{j=1}^N(v_j+2K)/\omega_1\right)
\label{sumeo2}
\end{equation}

{\bf Case 2 with $m_1$ odd}

\vspace{.1in}

The quasiperiodicity relation (\ref{om2qp}) also holds but now for
$m_1$ odd we have from (\ref{rdefs}) that $r_1/2$ is odd and thus
instead of (\ref{qpeo1}) we have
\begin{equation}
Q_{72oo}^{(1)}(v+\omega_1/2)=e^{N\pi i r_1r_2/4}RSQ_{72oo}^{(1)}(v)
\label{qpoo1}
\end{equation}
and therefore
\begin{equation}
[Q_{72oo}^{(1)}(v),RS]=0
\end{equation}
which has been directly shown in sec. 5.

It follows from the quasiperiodicity relations (\ref{qpoo1}) and
(\ref{qpeo2}) that the eigenvalues of $Q^{(1)}_{72oo}(v)$ are of the
form  (\ref{formeo}) where the sum rule 
(\ref{sumeo1}) is replaced by
\begin{equation}
e^{N \pi i r_1r_2/4}(-1)^{\nu_{RS}}=(-1)^\nu
\end{equation}
where $(-1)^{\nu_{RS}}$ are the eigenvalues of $RS$.

\subsection{Quasiperiodicity for $Q^{(2)}_{72ee}(v;t)$}

When $m_1$ and $m_2$ are both even we found that $Q^{(1)}_{72}$ does not
exist and that to solve the TQ equation we needed to use the matrix
$Q^{(2)}_{72}(v;t)$ constructed in sec. 6. This matrix exists only for
$N$ even and we recall that for $m_1$ and $m_2$ both even that $r_0$
is even. In fact we will see that for $t=(n+1/2)\eta$  there will 
be different cases for $r_0\equiv 0~({\rm mod 4})$ and 
$r_0\equiv 2~({\rm mod 4}).$  
 
We find from (\ref{SRm2}) and (\ref{hm1})-(\ref{tm2}) that
the quasi periodicity properties of $S^{(2)}_R(v)$ are
\begin{eqnarray}
&&S^{(2)}_R(\alpha,\beta)_{j,k}(v+\omega_1)
=(-\alpha)^{r_1}(-1)^{r_1r_2}S_R^{(2)}(\alpha,\beta)_{j,k}(v)\label{sqp21}\\
&&S^{(2)}_R(\alpha,\beta)_{k,k+1}(v+\omega_2)
=(-\alpha)^b(-1)^{ab}q'^{-1}e^{-2 \pi i(v-2k\eta-t-K)/\omega_1}
S^{(2)}_R(\alpha,\beta)_{k,k+1}(v)\label{sqp22}\\
&&S^{(2)}_R(\alpha,\beta)_{k+1,k}(v+\omega_2)
=(-\alpha)^b(-1)^{ab}q'^{-1}e^{-2 \pi i(v+2k\eta+t-K)/\omega_1}
S^{(2)}_R(\alpha,\beta)_{k+1,k}(v)\label{sqp23}
\end{eqnarray}
From (\ref{sqp21}) we find for all $t$ that $Q^{(2)}_{R}(v;t)$
has the periodicity property (recalling that $N$ is even)
\begin{equation}
Q^{(2)}_{R}(v+\omega_1;t)=(-S)^{r_1}Q^{(2)}_{R}(v;t)
\label{qrqpee1}\\
\end{equation}
However,the quasiperiodicity of $Q^{(2)}_R(v;t)$ under 
$v\rightarrow v+\omega_2$ is different for the two cases $t=n\eta$ and
$t=(n+1/2)\eta$ and will be treated separately.
\subsubsection{Quasiperiodicity for $t=n\eta$}  
When $t=n\eta$ the matrix $S^{(2)}(\alpha,\beta)(v+\omega_2)$ may be
written as
\begin{equation}
S^{(2)}_R(\alpha,\beta)(v+\omega_2)
=(-1)^{nr_0/2}(-\alpha)^b(-1)^{ab}q'^{-1}e^{-2\pi i(v-K)/\omega_1}
M^{(2;0)}S_R^{(2)}(\alpha,\beta)(v)M^{(2;0)-1}
\end{equation}
with $M^{(2;0)}$ given by  
\begin{equation}
M^{(2;0)}_{k,k'}=\delta_{k,k'}e^{-\pi i
  r_0k(k-1)/(2L)}(-1)^{nr_0k/2}e^{-\pi i nr_0k/(2L)}
 \end{equation}
and thus we obtain from (\ref{tqr0})
\begin{equation}
Q^{(2)}_{R}(v+\omega_2;n\eta)=(-S)^{b}q'^{-N}e^{-2\pi i
	N(v-K)/\omega_1}Q^{(2)}_{R}(v;t)\label{qrqpee2}
\end{equation}
Thus for $t=n\eta$ we find from the definition  (\ref{q72}) of
$Q^{(2)}_{72ee}(v;n\eta)$ that
\begin{eqnarray}
&&Q^{(2)}_{72ee}(v+\omega_1;n\eta)=(-S)^{r_1}Q^{(2)}_{72ee}(v;n\eta)
\label{qpee1}\\
&&Q^{(2)}_{72ee}(v+\omega_2;n\eta)=(-S)^{b}q'^{-N}e^{-2\pi i
	N(v-K)/\omega_1}Q^{(2)}_{72ee}(v;n\eta)\label{qpee2}
\end{eqnarray}
It follows from (\ref{rdefs}) that both $b$ and $r_1$
cannot both be even and thus $S$ appears in at least one of
(\ref{qpee1}) or (\ref{qpee2}) and therefore as
previously found in sec. 6  it follows that $Q^{(2)}_{72ee}(v;n\eta)$ 
commutes with the operator $S$. 

We further conclude from (\ref{qpee1}) and (\ref{qpee2}) that
the eigenvalues of $Q^{(2)}_{72ee}(v;n\eta)$ 
may be written in the form
\begin{equation}
Q^{(2)}_{72ee}(v;n\eta)={\cal K}{\rm exp}(-i\nu \pi
v/\omega_1)\prod_{j=1}^NH_m(v-v_j)
\label{eigenformee0}
\end{equation}
where the $N$ zeros $v_j$ are in the parallelogram (\ref{paroe}) and the sum 
rules  
\begin{eqnarray} 
&&e^{-i\pi \nu}=(-1)^{r_1(\nu_s-1)}\label{eesum1}\\
&&q'^{-\nu}{\rm exp}(2\pi i\sum_{j=1}^Nv_j/\omega_1)=
(-1)^{b(\nu_s-1)} \label{eesum2}
\end{eqnarray}
are satisfied. In particular if $r_1$ is even we see from
(\ref{eesum1}) that $\nu=0.$

\subsection{Quasiperiodicity for $t=(n+1/2)\eta$}

When $t=(n+1/2)\eta$ we first consider the case $m_2\equiv 0~({\rm
  mod}4)$. In this case 
$r_0\equiv 0~({\rm mod 4})$ and find from (\ref{sqp22}) and
(\ref{sqp23}) that
\begin{equation}
S_R^{(2)}(\alpha,\beta)(v+\omega_2;(n+1/2)\eta)
=(-1)^{r_0/4}(-\alpha)^b(-1)^{ab}q'^{-1}e^{-2\pi i(v-K)/\omega_1}
M^{(2;1)}S_R^{(2)}(\alpha,\beta)(v;(n+1/2)\eta)M^{(2;1)-1}
\label{simm21}
\end{equation}
where
\begin{equation}
M^{(2;1)}_{k,k'}=\delta_{k,k'}e^{-\pi i
  r_0k(k-1)/(2L)}(-1)^{r_0k/4}e^{-\pi i(2n+1)r_0k/(4L)}
\end{equation}
Thus from find from (\ref{tqr0}) that 
\begin{equation}
Q^{(2)}_R(v+\omega_2;(n+1/2)\eta)=(-S)^bq'^{-N}e^{-2\pi
  iN(v-K)/\omega_1}Q_R^{(2)}(v;(n+1/2)\eta)
\end{equation}
This is identical with (\ref{qrqpee2}) for $Q^{(2)}(v;n\eta)$ and thus we
conclude that for $r_0\equiv 0~({\rm mod 4})$ that
$Q^{(2)}_{72ee}(v;(n+1/2)\eta)$ has the same quasiperiodicity
properties (\ref{qpee1}),(\ref{qpee2}) 
and form of eigenvalues (\ref{eigenformee0}) 
as does $Q^{(2)}_{72ee}(v;n\eta)$ 

We next consider $m_2\equiv 2~({\rm mod}4)$ where $r_0\equiv 2~({\rm
  mod}4)$ $r_1$ is even and $r_2$ is odd.
In this case the similarity transformation 
in (\ref{simm21}) will not exist. The reason for this that in order for
(\ref{simm21}) to hold it was necessary that 
\begin{equation}
e^{\pi i r_0/4}=\pm 1
\end{equation}
which is not the case when $r_0\equiv 2~({\rm mod}4)$.
In this case the analogous argument shows that
$Q^{(2)}_{72ee}(v;(n+1/2)\eta)$ is quasi-periodic under $v\rightarrow
v+2\omega_2$ and thus has $2N$ zeros in the 
parallelogram $0,~\omega_1.~\omega_1+2\omega_2,2\omega_2$. 

However, these $2N$ zeros are not independent because there is an additional
quasiperiodicity under
$v\rightarrow v+\omega_2+\omega_1/2.$ To show this we use
 the relations which follow
from (\ref{more1})-(\ref{ThshC}) when $r_1$ is even and $r_2$ is odd
\begin{eqnarray}
 &&H_m(v+\omega_1/2+\omega_2)=-(-1)^{r_1/2+b}(-1)^{r_1r_2/4+ab}q'^{-1}
e^{-2\pi i(v-K)/\omega_1}\Theta_m(v)\label{klaus1}\\
 &&\Theta_m(v+\omega_1/2+\omega_2)=-(-1)^{r_1r_2/4+ab}q'^{-1}
e^{-2\pi i(v-K)/\omega_1}H_m(v) \label{klaus2}
\end{eqnarray}
and  find from (\ref{SRm2}) that
\begin{eqnarray}
&&S^{(2)}_R(\alpha,\beta)_{k,k+1}(v+\omega_1/2+\omega_2)
=(-\alpha)^{r_1/2+b}f(v)e^{\pi i/2}e^{2\pi i(t+2k\eta)/\omega_1}
S^{(2)}_R(-\alpha,\beta)_{k,k+1}(v)\label{sqp32}\\
&&S^{(2)}_R(\alpha,\beta)_{k+1,k}(v+\omega_1/2+\omega_2)
=(-\alpha)^{r_1/2+b}f(v)e^{-\pi i/2}
e^{-2\pi i(t+2k\eta)/\omega_1}
S^{(2)}_R(-\alpha,\beta)_{k+1,k}(v)\label{sqp33}\\
&&S^{(2)}(\alpha,\beta)_{1,L}(v+\omega_1/2+\omega_2)=(-\alpha)^{r_1/2+b}
f(v)e^{-\pi i/2}e^{-2\pi i(t+2L\eta)/\omega_1}
S^{(2)}_R(-\alpha,\beta)_{1,L}(v)\label{sqp34}\\
&&S^{(2)}_R(\alpha,\beta)_{L,1}(v+\omega_1/2+\omega_2)=(-\alpha)^{r_1/2+b}
f(v)e^{\pi i/2}e^{2\pi i(t+2L\eta)/\omega_1}
S^{(2)}_R(-\alpha,\beta)_{L,1}(v)
\label{sqp35}
\end{eqnarray}
where
\begin{equation}
f(v)=-i(-1)^{r_1r_2/4+ab}q'^{-1}e^{-2\pi i(v-K)/\omega_1}
\end{equation}
The expressions (\ref{sqp32})-(\ref{sqp35}) can be written as
\begin{equation}
M^{(2;2)}S^{(2)}_R(\alpha,\beta)(v+\omega_1/2+\omega_2)M^{(2;2)-1}
  =\epsilon (-\alpha)^{r_1/2+b}f(v)S^{(2)}_R(-\alpha,\beta)(v)
\label{sqp36}
\end{equation}
with $\epsilon=\pm1$ and
\begin{equation}
M^{(2;2)}_{k,k'}=m_k\delta_{k,k'}
\label{sqp37}
\end{equation}
where
\begin{equation}
m_k=(i\epsilon)^{k-1}e^{2\pi i (k-1)(t+k\eta)/\omega_1}
\end{equation}
and for consistency we need
\begin{equation}
(i\epsilon)^Le^{2\pi itL/\omega_1}e^{2\pi i L(L+1)\eta/\omega_1}=1
\label{consistancy}
\end{equation}
Using (\ref{etaom1}) and the fact that $r_0/2$  and $L$ are  
odd (\ref{consistancy}) reduces to
\begin{equation}
(i\epsilon)^Le^{\pi i tr_0/(2\eta)}=1
\end{equation}
which with an appropriate choice of $\epsilon=\pm 1$ 
is satisfied for $t=(n+1/2)\eta$ as desired.
Thus we obtain from (\ref{sqp36}) for $m_2\equiv 2~({\rm mod}4)$ and
$N$ even
\begin{equation}
Q^{(2)}_{72ee}(v+\omega_1/2+\omega_2;(n+1/2)\eta)
=(-1)^{N/2}S^{r_1/2+b}(-1)^{Nr_1r_2/4}q'^{-N}e^{-2\pi iN(v-K)/\omega_1}
RQ^{(2)}_{72ee}(v;(n+1/2)\eta)
\end{equation}

Finally we recall from (\ref{abdef}) that $b$ must be odd because $r_1$
is even and $r_2$ is odd and that because $r_0\equiv 2~({\rm mod}4)$
we have $r_1\equiv 2(0)~({\rm  mod}4)$ for $m_1\equiv 2(0)~({\rm mod
  4})$. Thus  we find that for $m_1\equiv 2~({\rm mod}4)$ and $m_2\equiv
2~({\rm mod}4)$ that
\begin{equation}
Q^{(2)}_{72ee}(v+\omega_1/2+\omega_2;(n+1/2)\eta)
=q'^{-N}e^{-2\pi iN(v-K)/\omega_1}RQ^{(2)}_{72ee}(v;(n+1/2)\eta)
\label{finalqpm1o}
\end{equation}
from which it follows in agreement with (\ref{c4c}) that
$Q^{(2)}_{72ee}(v;(n+1/2)\eta)$ commutes with $R.$
For $m_2\equiv 2~({\rm mod}4)$ 
and $m_1\equiv 0~({\rm mod} 4)$
\begin{equation}
Q^{(2)}_{72ee}(v+\omega_1/2+\omega_2;(n+1/2)\eta)
=(-1)^{N/2}q'^{-N}e^{-2\pi iN(v-K)/\omega_1}RSQ^{(2)}_{72ee}(v;(n+1/2)\eta)
\label{finalqpm1e}
\end{equation}
from which it follows in agreement with (\ref{c4d}) that
$Q^{(2)}_{72ee}(v;(n+1/2)\eta)$ commutes with $RS$

It follows from (\ref{finalqpm1o}) and (\ref{finalqpm1e}) that for
$m_2\equiv 2~({\rm mod}4)$ there are $N$ zeros in the fundamental
parallelogram
\begin{equation}
0,~\omega_1,~3\omega_1/2+\omega_2,~\omega_1/2+\omega_2
\end{equation}

For $m_1\equiv 2~({\rm mod}4)$ and $m_2\equiv
2~({\rm mod}4)$ it follows from (\ref{finalqpm1o}) that the eigenvalues
of $Q^{(2)}_{72ee}(v;(n+1/2)\eta)$ are of the form
\begin{eqnarray}
&&Q^{(2)}_{72ee}(v;(n+1/2)\eta)\nonumber\\
&&={\cal K}
\prod_{j=1}^NH_m(\frac{v-v_j}{2})
H_m(\frac{v-v_j+\omega_1/2+\omega_2}{2})
H_m(\frac{v-v_j+\omega_1+2\omega_2}{2})
H_m(\frac{v-v_j+3\omega_1/2+3\omega_2}{2})
\label{formeeetao2}
\end{eqnarray}
which as required by the condition (\ref{qrqpee1}) 
is periodic in $v\rightarrow v+\omega_1$ and satisfies the
quasiperiodicity condition (\ref{finalqpm1o}) with the sum rule 
\begin{equation}
(-1)^{\nu_R}=q'^{-3N}e^{2\pi i\sum_{j=1}^N(v_j+K)/\omega_1}
=q'^{-3N-Nr_2}e^{2\pi i\sum_{j=1}^Nv_j/\omega_1}
\end{equation}

For $m_1\equiv 0~({\rm mod}4)$ and $m_2\equiv
2~({\rm mod}4)$ it follows from (\ref{finalqpm1e}) that the eigenvalues
of $Q^{(2)}_{72ee}(v;(n+1/2)\eta)$ are of the form (\ref{formeeetao2})
with the sum rule
\begin{equation}
(-1)^{N/2}(-1)^{\nu_{RS}}=q'^{-3N}e^{2\pi i\sum_{j=1}^N(v_j+K)/\omega_1}
=q'^{-3N-Nr_2}e^{2\pi i\sum_{j=1}^Nv_j/\omega_1}
\end{equation}

\section{Conclusion}

In this paper we have extended the construction of the matrices
$Q_{72}(v)$  that solve the TQ equation (\ref{tq1})-(\ref{tq3}) 
which were introduced in \cite{bax72} to solve the eight vertex model
in the elliptic root of unity case (\ref{tq4}) from $m_2=0$ to
$m_2\neq 0$ and have found that the constructions depend on the
parities of $m_1$ and $m_2$. In all cases we have examined the
matrices $Q_{72}(v)$ are nondegenerate. 

We have found that in all cases the
matrix $Q_{72}(v)$ defined at roots of unity (\ref{tq4}) 
commutes with only one of the three matrices $S,~R$
and $RS$ which are the discrete symmetries of the transfer matrix
$T(v)$ of the eight-vertex model. This is in contrast with the matrix
$Q_{73}(v)$ of \cite{bax731} which commutes with all three of the discrete
symmetry operators. This failure of $Q_{72}(v)$ to commute with one
of the discrete symmetry operators of $T(v)$ when combined with the
nondegeneracy of $Q_{72}(v)$  provides  an explanation
of the existence of  some eigenvalues of $T(v)$ which are at least
doubly degenerate. This  mechanism is not possible for $Q_{73}(v)$ at
roots of unity and is an indication that thee is a sense in which
$Q_{72}(v)$ contains information which is lacking in $Q_{73}(v)$.

Perhaps the most novel feature of our results is that in the case
where both $m_1$ and $m_2$ are even that there are different cases
depending on whether or not $m_1$ and $m_2$ are divisible by four
and that for $m_2\equiv 2~({\rm mod}4)$ there exist two matrices
with different commutation properties with the discrete symmetry
operators. 
These matrices map a degenerate subspace of $T$ on another degenerate subspace
with opposite eigenvalue of the discrete symmetry operator thereby 
doubling the size of degenerate muliplets of T.

We finally note that even though there are cases 
where $Q_R^{(1)}(v)$ or  $Q_R^{(2)}(v)$ obey the interchange relation
(\ref{commlra})
with all four of the operators $I,S,R$ and $RS$ there is in fact no
matrix $Q_{72}(v)$ which shares with $Q_{73}(v)$ the property
of commuting with all three operators $R,S$ and $RS.$
This would happen for  $Q_R^{(1)}(v)$ if $m_1$ and $m_2$ are 
even\footnote{See Table 1.} and for $Q_R^{(2)}(v;t)$
if $m_1\equiv 2~({\rm mod}4)$ and $m_2\equiv 0~({\rm mod}4)$ 
and $t=(n+1/2)\eta$ \footnote{See Table 3.}, but in these cases 
$Q_R(v)$ is singular and thus the construction (\ref{q72}) of $Q_{72}(v)$
cannot be made.

The various new properties of $Q_{72}(v)$ found in this paper for $m_2\neq 0$ 
must contain useful
information about the still undetermined symmetry algebra of the eight
vertex model at elliptic roots of unity.

\app{Properties of
 the modified theta functions}

The functions $H(v)$ and $\Theta(v)$ defined by (\ref{hdef}) and
 (\ref{thetadef}) have the following
well known properties
\begin{eqnarray}
&&H(-v)=-H(v),
\hspace{1in}\Theta(-v)=\Theta(v)\label{napp1}\\
&&H(v+2nK)=(-)^nH(v), 
\hspace{.5in}
\Theta(v+2nK)=\Theta(v)\label{app3}\\
&&H(v+2inK')=(-1)^nq^{-n^2}e^{-n\pi i v/K}H(v)\label{napp2}\\
&&\Theta(v+2inK')=(-1)^nq^{-n^2}e^{-n\pi i v/K}\Theta(v)\label{napp4}\\
&&\Theta(v+iK')=iq^{-1/4}e^{-\frac{\pi i v}{2K}}H(v),
\hspace{.5in}
H(v+iK')=iq^{-1/4}e^{-\frac{\pi i v}{2K}}\Theta(v)\label{app6}
\end{eqnarray}

It follows immediately from (\ref{napp1})-(\ref{app6}) that
the modified theta functions  $H_m(v)$ and $\Theta_m(v)$ defined by
(\ref{Hm})  have the properties
\begin{equation}
H_m(2K-v) = H_m(v),
\hspace{0.5 in}
\Theta_m(2K-v) = \Theta_m(v)
\label{help1}
\end{equation}
\begin{equation}
H_m(-v) = -\exp\left(\frac{i\pi m_2 v}{2L\eta}\right)H_m(v)
\hspace{0.5 in}
\Theta_m(-v) = \exp\left(\frac{i\pi m_2 v}{2L\eta}\right)\Theta_m(v)
\label{Hminus}
\end{equation}
\begin{equation}
H_m(u+2rK+2isK')
=(-1)^{r}(-1)^{rs}H_m(u){\rm exp}\{(\pi i(rm_2/2-sm_1)[u+(r-1)K+isK']/(L\eta)\}
\label{more1}
\end{equation}
and
\begin{equation}
\Theta_m(u+2rK+2isK')
=(-1)^{rs}\Theta_m(u){\rm exp}\{(\pi i(rm_2/2-sm_1)[u+(r-1)K+isK']/(L\eta)\}
\label{more2}
\end{equation}
\begin{equation}
\Theta_m(v+iK') 
= i q^{-1/4}\exp\left(\frac{-i\pi m_1v}{2L\eta}\right)C
H_m(v),~~~ H_m(v+iK') 
= i q^{-1/4}\exp\left(\frac{-i\pi m_1v}{2L\eta}\right)C\Theta_m(v)
\label{Thsh}
\end{equation}
where
\begin{equation}
C = \exp\left(\frac{\pi m_2K'}{8KL\eta}(2K-iK')\right)
\label{ThshC}
\end{equation}
For convenience we note the special cases
\begin{equation}
H_m(v+2L\eta) = (-1)^{m_1}i^{m_1m_2}
\left\{\begin{array}{ll}
H_m(v) & \mbox{ if $m_2$=even}\\
\Theta_m(v) & \mbox{ if $m_2$=odd}\\
\end{array}\right. \
\label{Hm2Leta}
\end{equation}
\begin{equation}
\Theta_m(v+2L\eta) = i^{m_1m_2}
\left\{\begin{array}{ll}
\Theta_m(v) & \mbox{ if $m_2$=even}\\
H_m(v) & \mbox{ if $m_2$=odd}\\
\end{array}\right. \
\label{Tm2Leta}
\end{equation}
The properties (\ref{Hm2Leta}),(\ref{Tm2Leta}),
(\ref{hm1})-(\ref{tm2}) follow immediately from (\ref{more1})-(\ref{ThshC}).

The functions $H_m(v)$ and $\Theta_m(v)$ satisfy  the identities
\begin{eqnarray}
&&H_m(u)H_m(v)H_m(w)H_m(u+v+w)+\Theta_m(u)\Theta_m(v)\Theta_m(w)
\Theta_m(u+v+w)\nonumber\\
&&=\Theta_m(0)\Theta_m(u+v)\Theta_m(u+w)\Theta_m(v+w)
\label{htident1}
\end{eqnarray}
\begin{eqnarray}
&&H_m(u)H_m(v)\Theta_m(w)\Theta_m(u+v+w)+\Theta_m(u)\Theta_m(v)H_m(w)H_m(u+v+w)
\nonumber\\
&&=\Theta_m(0)\Theta_m(u+v)H_m(u+w)H_m(v+w)
\label{htident2}
\end{eqnarray}
\app{Modified theta functions and Jacobi theta functions.}
We write the modified theta functions
\begin{equation}
H_m(u) = \exp\left( \frac{i\pi m_2(u-K)^2}{8KL\eta}\right)H(u)
\hspace{0.5 in}
\Theta_m(u) = \exp\left(\frac{i\pi m_2(u-K)^2}{8KL\eta}\right)\Theta(u)
\label{Thetamod0}
\end{equation}
in terms of theta functions with characteristics defined as
\begin{equation}
\Theta_{\epsilon\,\epsilon'}
= \sum_{ -\infty}^{\infty}q^{(n+\epsilon/2)^2}\exp(2\pi i(n+\epsilon/2)(z+\epsilon'/2))
\label{Thetachar}
\end{equation}
\begin{equation}
q = \exp(i\pi \tau)
\end{equation}  
To cancel a common divisor of $m_1$ and $m_2$ in
\begin{equation}
2L\eta = 2m_1K+im_2K'
\end{equation}
we define
\begin{equation}
(2m_1,m_2) = r_0
\hspace{0.5 in} 2m_1 = r_0 r_1 
\hspace{0.3 in} m_2 = r_0 r_2
\label{r1r2} 
\end{equation}
\begin{equation}
2L\eta = r_0(r_1K+ir_2K')
\end{equation}
It follows
\begin{equation}
H_m(u) = -\exp\left(\frac{i\pi r_2w^2}{r_1+r_2\tau}\right)
\Theta_{11}(w+1/2)
\end{equation}
\begin{equation}
\Theta_m(u) = ~~\exp\left(\frac{i\pi r_2w^2}{r_1+r_2\tau}\right)
\Theta_{01}(w+1/2)
\end{equation}
where 
\begin{equation}
w = u/2K-1/2
\end{equation}
or after a shift of the argument
\begin{equation}
H_m(u) = \exp\left(\frac{i\pi r_2w^2}{r_1+r_2\tau}\right)
\Theta_{10}(w)
\label{Hmod}
\end{equation}
\begin{equation}
\Theta_m(u) = \exp\left(\frac{i\pi r_2w^2}{r_1+r_2\tau}\right)
\Theta_{00}(w)
\label{Thetamod2}
\end{equation}
The second step is to use the functional relation of theta functions.
From \cite{FarKra}
\begin{equation}
\exp\left(\frac{i\pi cw^2}{c\tau+d}\right)
\Theta_{\tilde{\epsilon},\tilde{\epsilon}'}(w)=
\kappa^{-1}(\epsilon,\epsilon',\gamma)(c\tau+d)^{-1/2}
\Theta_{\epsilon,\epsilon'}
\left(\frac{w}{c\tau+d},\frac{a\tau +b}{c\tau +d}\right)
\label{functrel}
\end{equation}
where
\begin{equation}
\gamma = \left( 
\begin{array}{cc}
a&b\\
c&d\\
\end{array}
\right)
\hspace{0.4 in}ad-bc=1
\end{equation} 
and
\begin{equation}
\tilde{\epsilon} = a\epsilon + c \epsilon'-ac 
\hspace{0.5 in}
\tilde{\epsilon}' = b\epsilon + d \epsilon'+bd 
\end{equation}
We use (\ref{functrel}) to derive from (\ref{Hmod}) and (\ref{Thetamod2}) for $c=r_2$ and $d=r_1$
\begin{equation}
H_m(u) = \kappa(\epsilon_1,\epsilon_1',\gamma)^{-1}(r_1+r_2\tau)^{-1/2}
\Theta_{\epsilon_1,\epsilon_1'}\left(\frac{w}{r_1+r_2\tau},\tau'\right)
\label{Hma}
\end{equation}
\begin{equation}
\Theta_m(u) = \kappa(\epsilon_2,\epsilon_2',\gamma)^{-1}(r_1+r_2\tau)^{-1/2}
\Theta_{\epsilon_2,\epsilon_2'}\left(\frac{w}{r_1+r_2\tau},\tau'\right)
\label{Tm}
\end{equation}
\begin{equation}
\epsilon_1 = ar_1(r_1+r_2)
\hspace{0.5 in}
\epsilon_1' = -b(1+a(r_1+r_2))
\end{equation}
\begin{equation}
\epsilon_2 = r_1r_2(a+b)
\hspace{0.5 in}
\epsilon_2' = -ab(r_1+r_2)
\end{equation}
where 
\begin{equation}
\tau' = \frac{a\tau+b}{r_2\tau+r_1}
\hspace{0.3 in}
w=\frac{u-K}{2K}
\end{equation}
and where the integers $a,b$ are solutions of
\begin{equation}
ar_1-br_2=1
\end{equation}
The indices $\epsilon_1,\cdots \epsilon_2'$ can be shifted to $0,1$ such that the theta functions on the
right hand sides of (\ref{Hma}), \ref{Tm}) become
\[
\Theta_{00},\Theta_{01},\Theta_{10},\Theta_{11} \equiv  \Theta_1,\Theta,H_1,H.
\]
It follows from equs. (\ref{Hma}) and (\ref{Tm}) that the period and quasiperiod of $H_m(u)$ and $\Theta_m(u)$ are 
\begin{equation}
\omega_1 = 2(r_1K+ir_2K')  \hspace{0.5 in} \omega_2 = 2(bK+iaK') 
\end{equation}
The coefficient $\kappa(\epsilon,\epsilon',\gamma)$ is not used in this paper. 
It is an eighth root of unity and its dependence on $\epsilon,\epsilon'$ is given by \cite{FarKra}.
\begin{equation}
\kappa(\epsilon,\epsilon',\gamma) = \kappa(0,0,\gamma) 
\exp\left(\frac{-i\pi}{4}(\epsilon^2ab+\epsilon'^2cd + 2\epsilon\epsilon'bc+2(a\epsilon+c\epsilon')bd)\right)
\end{equation}

\app{The equation for $Q^{(2)}_{R}(v;t)$}

We establish here the relation (\ref{2tq1}).
Using the method of ref. \cite{bax72} and \cite{klaus} we write
\begin{equation}
T(v)Q^{(2)}_R(v;t)={\rm Tr}A(\alpha_1,\beta_1)\cdots A(\alpha_N,\beta_N)
+{\rm Tr}B(\alpha_1,\beta_1)\cdots B(\alpha_N,\beta_N)  
\label{2tq3}
\end{equation}
where $A(\alpha,\beta)$ and $B(\alpha,\beta)$ are $2L\times 2L$ matrices given
by
\begin{equation}
\left(\begin{array}{cc}
A(\alpha,\beta)&0\\
C(\alpha,\beta)&B(\alpha,\beta)
\end{array}\right)
=\left(\begin{array}{cc}
I&-P\\
0&I 
\end{array}\right)U(\alpha,\beta)
\left(\begin{array}{cc}
I&P\\
0&I
\end{array}\right)
\label{2tq4}
\end{equation}
with
\begin{equation}
P_{m,n}=p_n\delta_{m,n}
\label{2tq5}
\end{equation}
and $p_n$ given by (\ref{pn2})
and where
\begin{eqnarray}
&&U(+,\beta)=\left(
\begin{array}{cc}
aS^{(2)}_R(+,\beta)&dS^{(2)}(-,\beta)\\
cS^{(2)}_R(-,\beta)&bS^{(2)}_R(+,\beta)
\end{array}\right)\\
&&U(-,\beta)=\left(
\begin{array}{cc}
bS^{(2)}_R(-,\beta)&cS^{(2)}(+,\beta)\\
dS^{(2)}_R(+,\beta)&aS^{(2)}_R(-,\beta)
\end{array}\right)
\end{eqnarray}

Then from (\ref{2tq4}) and (\ref{SRm2}) we use
\begin{equation}
Hm(-v)/\Theta_m(-v)=-H_m(v)/\Theta_m(v)
\end{equation}
and the identities (\ref{htident1}), (\ref{htident2}) to
find
\begin{equation}
\label{a1}
A(+,\beta)_{k,k+1} = 
 -\tau_{\beta,-k}
\frac{\Theta_m(-(2k+1)\eta-t)}
{\Theta_m(-(2k-1)\eta-t)}h(v-\eta)H_m(v-2k\eta+2\eta-t) 
\end{equation}
\begin{equation}
A(+,\beta)_{k+1,k} = 
 \tau_{\beta,k}\frac{\Theta((2k-1)\eta+t)}{\Theta((2k+1)\eta+t)}
h(v-\eta)H_m(v+2k\eta+2\eta+t) 
\label{a2}
\end{equation}
\begin{equation}
\label{a3}
A(-,\beta)_{k,k+1} = 
  \tau_{\beta,-k}\frac{\Theta_m(-(2k+1)\eta-t)}
{\Theta_m(-(2k-1)\eta-t)}h(v-\eta)\Theta_m(v-2k\eta+2\eta-t) 
\end{equation}
\begin{equation}
A(-,\beta)_{k+1,k} = 
\tau_{\beta,k}\frac{\Theta_m((2k-1)\eta+t)}{\Theta_m((2k+1)\eta+t)}
h(v-\eta)\Theta_m(v+2k\eta+2\eta+t) 
\label{a4}
\end{equation}
\begin{equation}
\label{b1}
B(+,\beta)_{k,k+1} = 
 -\tau_{\beta,-k}\frac{\Theta_m(-(2k-1)\eta-t)}{\Theta_m(-(2k+1)\eta-t)}
h(v+\eta)H_m(v-2k\eta-2\eta-t) 
\end{equation}
\begin{equation}
\label{b2}
B(+,\beta)_{k+1,k} = 
\tau_{\beta,k}\frac{\Theta_m((2k+1)\eta+t)}{\Theta_m((2k-1)\eta+t)}
h(v+\eta)H_m(v+2k\eta-2\eta+t) 
\end{equation}
\begin{equation}
\label{b3}
B(-,\beta)_{k,k+1} = 
\tau_{\beta,-k}\frac{\Theta_m(-(2k-1)\eta-t)}{\Theta_m(-(2k+1)\eta-t)}
h(v+\eta)\Theta_m(v-2k\eta-2\eta-t)
 \end{equation}
\begin{equation}
B(-,\beta)_{k+1,k} = 
\tau_{\beta,k}\frac{\Theta_m((2k+1)\eta+t)}{\Theta_m((2k-1)\eta+t)}
h(v+\eta)\Theta_m(v+2k\eta-2\eta+t) 
\label{b4}
\end{equation}

From (\ref{Hminus}) we have
\begin{equation}
\frac{\Theta_m(-(2k+1)\eta-t)}{\Theta_m(-(2k-1)\eta-t)}
={\rm exp}(\pi im_2/L)\frac{\Theta_m((2k+1)\eta+t)}{\Theta_m((2k-1)\eta+t)}
\end{equation}
and thus with the definition (\ref{2tq2}) of $\omega$ and 
\begin{equation}
f_k = \omega  \frac{\Theta((2k+1)\eta+t)}{\Theta((2k-1)\eta+t)}
\end{equation}
we may write (\ref{a1})-(\ref{b4}) as
\begin{equation}
A(+,\beta)_{k,k+1}(v) = \omega f_k~~ h(v-\eta) S_R(+,\beta)_{k,k+1}(v+2\eta)
\label{a11}
\end{equation}
\begin{equation}
A(+,\beta)_{k+1,k}(v) = \omega f_k^{-1} h(v-\eta) S_R(+,\beta)_{k+1,k}(v+2\eta) 
\label{a12}
\end{equation}
\begin{equation}
A(-,\beta)_{k,k+1}(v) = \omega f_k~~ h(v-\eta) S_R(-,\beta)_{k,k+1}(v+2\eta) 
\label{a13}
\end{equation}
\begin{equation}
A(-,\beta)_{k+1,k}(v) = \omega f_k^{-1} h(v-\eta) S_R(-,\beta)_{k+1,k}(v+2\eta)
 \label{a14}
\end{equation}
\begin{equation}
B(+,\beta)_{k,k+1}(v) = \omega^{-1} f_k^{-1} h(v+\eta) S_R(+,\beta)_{k,k+1}(v-2\eta) 
\label{b11}
\end{equation}
\begin{equation}
B(+,\beta)_{k+1,k}(v) 
= \omega^{-1} f_k~~ h(v+\eta) S_R(+,\beta)_{k+1,k}(v-2\eta) 
\label{b12}
\end{equation}
\begin{equation}
B(-,\beta)_{k,k+1}(v) = \omega^{-1} f_k^{-1} h(v+\eta) S_R(-,\beta)_{k,k+1}(v-2\eta) 
\label{b13}
\end{equation}
\begin{equation}
B(-,\beta)_{k+1,k}(v) = \omega^{-1} f_k~~ 
h(v+\eta) S_R(-,\beta)_{k+1,k}(v-2\eta) 
\label{b14}
\end{equation}

The $TQ_R$ equation (\ref{2tq3}) will be obtained if the factors of
$f_k$ can be removed by a diagonal similarity transformation. 
\begin{equation}
S_AA(\alpha,\beta)S^{-1}_A
\end{equation}
with
\begin{equation}
S_{A;k,k'}=s_k\delta_{k,k'}
\end{equation}
this is accomplished for the elements $A_{k,k+1}(\alpha,\beta)$ and 
$A_{k+1,k}(\alpha,\beta)$ with $1\leq k, \leq L-1$ 
\begin{equation}
\frac{s_kf_k}{s_{k+1}}=\frac {s_{k+1}}{s_kf_k}=\frac{s_1}{s_Lf_L}
=\frac{s_Lf_L}{s_1}~~~{\rm for}~~1\leq k
\leq L-1
\label{skdef}
\end{equation}
From the first equation in (\ref{skdef}) 
we have
\begin{equation}
\frac{s_kf_k}{s_{k+1}}=\pm 1
\label{sim1}
\end{equation}
where the choice $\pm 1$ is still to be determined and from
(\ref{sim1}) we have
\begin{equation}
s_k=(\pm 1)^{k-1}s_1\omega^{k-1}\frac{\Theta_m[(2k-1)\eta+t]}{\Theta_m
  (\eta+t)}.
\label{sim2}
\end{equation}
The remaining equations in (\ref{sim1}) will hold if 
\begin{equation}
\frac{s_Lf_L}{s_1}=\pm1
\label{sim3}
\end{equation}
and using (\ref{sim2}) we obtain
\begin{equation}
(\pm 1)^L\omega^L\frac{\theta_m[(2L+1)\eta+t]}{\theta_m(\eta+t)}=1
\label{sim4}
\end{equation}
which if we further use (\ref{Tm2Leta}) restricted to the present case where
$m_1$ and $m_2$ are even and $L$ is odd determines that the factor
$\pm 1$ is
\begin{equation}
\pm 1=(-1)^{m_2/2}
\end{equation}

An identical computation holds for the matrices $B(\alpha,\beta)$ and thus
(recalling the $N$ is even) we have proven that (\ref{2tq1}) holds.


\begin{thebibliography}{99}



\bibitem{bax72} R.J. Baxter, Partition function of the eight vertex
model, Ann. Phys. 70 (1972) 193-228.


\bibitem{bax731}R.J. Baxter,
Eight--vertex model in lattice statistics and one--dimensional
anisotropic Heisenberg chain I: Some fundamental eigenvectors,
Ann. Phys. 76 (1973) 1-24.

\bibitem{bax732}R.J. Baxter,
Eight--vertex model in lattice statistics and one--dimensional
anisotropic Heisenberg chain II: Equivalence to a generalized
Ice-type lattice model, Ann. Phys. 76 (1973) 25-47.


\bibitem{bax733}R.J. Baxter, Eight--vertex model in
lattice statistics and one--dimensional
anisotropic Heisenberg chain III: Eigenvectors of the  transfer matrix
and the  Hamiltonian,
Ann. Phys. 76 (1973)48-71.


\bibitem{bax02}R. J. Baxter, On the completeness of the Bethe
Ansatz for the six and eight-vertex models
J. Stat. Phys. 108 (2002) 1-48.

\bibitem{tf} L.A. Takhtadzhan and L.D. Faddeev, The quantum method of 
the inverse
problem and the Heisenberg XYZ model, Russian Math. Surveys 34:5 (1779)
11-68; translated from Uspekhi Mat. Nauk 34:5 (1979) 12-63.

\bibitem{Zab} A. Zabrodin, Commuting Difference 
Operators with elliptic coefficients from
Baxter's vacuum vectors,
J. Phys. A:Mat.Gen. 33,(2000) 3825-3850.


\bibitem{deg1} T. Deguchi, Construction of some missing
eigenvectors of the XYZ spin chain at discrete coupling constants
and the exponentially large spectral degeneracy of the transfer matrix,
J.Phys. A 35 (2002) 879-895.

\bibitem{deg2} T. Deguchi, The 8V CSOS model and the $sl_2$
loop algebra symmetry of the six-vertex model at roots of unity,
Int. J. Mod. Phys. B 16 (2002) 1899-1905.


\bibitem{baz1} V.V. Bazhanov and V.V. Mangazeev, Eight vertex model and
  non-stationary Lame equation, J. Phys. A:Math.Gen. 38 (2005) L145-L153.

\bibitem{baz2} V.V. Bazhanov and V.V. Mangazeev, Eight vertex model
  and Painlev{\'e} VI, J. Phys. A: Math.Gen. 39 (2006) 12235-12243.

\bibitem{baz3} V.V. Bazhanov and V.V. Mangazeev, Analytic theory of
  the eight vertex model, Nucl. Phys. B 775 (2007) 225-282.




\bibitem{roan} S-S. Roan, The Q-operator and functional relations
  of the eight-vertex model at root-of-unity $\eta=2mK/N$ for odd $N$,
J. Phys. A:Math.Theor. 40 (2007), 11019-11044. 


\bibitem{fm1} K.Fabricius and B.M.McCoy, New developements in the eight
  vertex model, J. Stat. Phys. 111 (2003) 323-337.


\bibitem{fm2}  K. Fabricius and B.M. McCoy, New developments in the eightvertex
model II. Chains of odd length, J. Stat. Phys. 120 (2005) 37-70.

\bibitem{klaus} K. Fabricius, A new Q matrix in the eightvertex model,
J. Phys. A: Math. Theor. 40 (2007) 4075-4086.
 
\bibitem{strog1} Yu. Stroganov, The importance of being odd,
  J. Phys. A: Math.Gen. 34 (2001) L179-L185.

\bibitem{strog2} A.V. Razumov and Yu. G. Stroganov, Spin chains and
  combinatorics, J. Phys. A: Math.Gen. 34 (2001) 3185-3190.

\bibitem{murray1} M.T. Batchelor, J.De Gier and B. Nienhuis, The
  quantum symmetric XXZ chain at $\Delta=-1/2$, alternating-sign
  matrices and plane partitions, J. Phys. A 34 (2001) L265-L270.

\bibitem{murray} J. de Gier, M.T. Batchelor, B. Nienhuis, and
  S. Mitra, The XXZ spin chain at $\Delta=-1/2$; Bethe roots,
  symmetric functions and determinants, J. Math. Phys. 43 (2002) 4135-4146.

\bibitem{baxbook} R. J. Baxter, {\it Exactly Solved Models in Statistical Mechanics}, Academic Press.
London (1982).



\bibitem{FarKra} H. Farkas and I. Kra, Theta constants,
Riemann surfaces and the modular group, Graduate Studies in
Mathematics, Vol 37 (Am. Math. Soc. Providence Rhode Island) 2001.






\end{thebibliography}
\end{document}